\documentclass[useAMS,usenatbib,onecolumn]{mn2e}
\usepackage{graphicx}

\title[Properties of Giant Molecular Clouds in the LMC]{Physical
  Properties of Giant Molecular Clouds in the Large Magellanic Cloud}
\author[A. Hughes et al.]{A.~Hughes$^{1,2}$\thanks{E-mail:
    ahughes@astro.swin.edu.au}, T.~Wong$^{3}$, J.~Ott$^{4}$,
  E.~Muller$^{5}$, J.~L.~Pineda$^{6}$\thanks{NASA Postdoctoral Program
    Fellow}, Y.~Mizuno$^{5}$, \newauthor J.-P.~Bernard$^{7}$,
  D.~Paradis$^{8}$, S.~Maddison$^{1}$, W.~T.~Reach$^{8}$,
  L.~Staveley-Smith$^{9}$, \newauthor A.~Kawamura$^{5}$,
  M.~Meixner$^{10,11}$, S.~Kim$^{12}$, T.~Onishi$^{5,13}$,
  N.~Mizuno$^{5,14}$, Y.~Fukui$^{5}$\\ $^1$Centre for Supercomputing
  and Astrophysics, Swinburne University of Technology, Hawthorn VIC
  3122, Australia \\ $^2$CSIRO Australia Telescope National Facility,
  PO Box 76, Epping NSW 1710, Australia \\ $^3$Astronomy Department,
  University of Illinois, 1002 W. Green St, Urbana, IL 61801,
  USA\\ $^4$National Radio Astronomy Observatory, P.O. Box O, 1003
  Lopezville Road, Socorro, NM 87801, USA\\ $^5$Department of
  Astrophysics, Nagoya University, Furo-cho, Chikusa-ku, Nagoya
  464-8602, Japan\\ $^6$Jet Propulsion Laboratory, California
  Institute of Technology, 4800 Oak Grove Drive, Pasadena, CA
  91109-8099, USA\\ $^7$Centre d'Etude Spatiale des Rayonnements,
  Universite Paul Sabatier, 9 Av du Colonel Roche, BP 44346, Toulouse,
  France\\ $^8${\it Spitzer} Science Center, Caltech, MS220-6,
  Pasadena, CA 91125, USA\\ $^9$International Centre for Radio
  Astronomy Research, M468, University of Western Australia, 35
  Stirling Hwy, Crawley, WA 6009 \\ $^{10}$Space Telescope Science
  Institute, 3700 San Martin Drive, Baltimore, MD 21218,
  USA\\ $^{11}$Radio and Geoastronomy Division, Harvard-Smithsonian
  for Astrophysics, 60 Garden St, MS 42, Cambridge, MA 02138-1516, USA
  \\ $^{12}$Department of Astronomy and Space Science, Sejong
  University, KwangJin-gu, KunJa-dong 98, Seoul 143-747,
  Korea\\ $^{13}$Department of Physical Science, Osaka Prefecture
  University, Gakeun 1-1, Sakai, Osaka 599-8531, Japan\\ $^{14}$ALMA-J
  Project Office, National Astronomical Observatory of Japan, 2-21-1
  Osawa, Mitaka, Tokyo 181-8588, Japan\\}

\newcommand{\ea}{\emph{et al.}}
\def\ts   {\thinspace}
\def\kms  {\ifmmode{{\rm \ts km\ts s}^{-1}}\else{\ts km\ts s$^{-1}$}\fi}
\def\kkms  {\ifmmode{{\rm \ts K\ts km\ts s}^{-1}}\else{\ts K\ts km\ts s$^{-1}$}\fi}
\def\xco   {\ifmmode{X_{\rm CO}}\else{$X_{\rm CO}$}\fi}
\def\xcou  {\ifmmode{{\rm \ts cm^{-2}\ts (K\ts km\ts s^{-1})}^{-1}}\else{\ts cm$^{-2}$\ts (K\ts km\ts s$^{-1}$)$^{-1}$}\fi}
\def\phu  {\ifmmode{{\rm \ts 10^{4}\ts k_{B}\ts cm^{-3}\ts K}}\else{\ts 10$^{4}$\ts k$_{B}$\ts cm$^{-3}$\ts K}\fi}
\def\pcmsq  {\ifmmode{{\rm \ts cm}^{-2}}\else{\ts cm$^{-2}$}\fi}
\def\cc  {\ifmmode{{\rm \ts cm}^{-2}}\else{\ts cm$^{-2}$}\fi}
\def\ccc  {\ifmmode{{\rm \ts cm}^{-3}}\else{\ts cm$^{-3}$}\fi}
\def\msol   {\ifmmode{{\rm M}_{\odot}}\else{M$_{\odot}$}\fi}
\def\mpcsq   {\ifmmode{{\rm M}_{\odot}\ts {\rm pc}^{-2}}\else{M$_{\odot}$}\ts pc$^{-2}$\fi}
\def\aco {\ifmmode{^{12}{\rm CO}(J=1\to0)}\else{$^{12}{\rm
CO}(J=1\to0)$}\fi}
\def\bco {\ifmmode{^{12}{\rm CO}(J=2\to1)}\else{$^{12}{\rm
CO}(J=2\to1)$}\fi}
\def\hi  {\ifmmode{{\rm H}{\rm \scriptsize I}}\else{H\ts {\scriptsize I}}\fi}
\def\hii  {\ifmmode{{\rm H}{\rm \small II}}\else{H\ts {\scriptsize II}}\fi}
\def\nh  {\ifmmode{N(\hi)}\else{$N$(\hi)}\fi}
\def\hh   {\ifmmode{{\rm H}_2}\else{H$_2$}\fi}
\def\ha   {\ifmmode{{\rm H}\alpha)}\else{H$\alpha$}\fi}

\begin{document}
\date{Typeset \today; Received / Accepted}
\maketitle
\label{firstpage}

\begin{abstract}
\label{sect: abstract}

\noindent The Magellanic Mopra Assessment (MAGMA) is a high angular
resolution \aco\ mapping survey of giant molecular clouds (GMCs) in
the Large and Small Magellanic Clouds using the Mopra Telescope.  Here
we report on the basic physical properties of 125 GMCs in the Large
Magellanic Cloud (LMC) that have been surveyed to date. The observed
clouds exhibit scaling relations that are similar to those determined
for Galactic GMCs, although LMC clouds have narrower linewidths and
lower CO luminosities than Galactic clouds of a similar size. The
average mass surface density of the LMC clouds is 50~\mpcsq,
approximately half that of GMCs in the inner Milky Way. We compare the
properties of GMCs with and without signs of massive star formation,
finding that non-star-forming GMCs have lower peak CO brightness than
star-forming GMCs. The average CO-to-\hh\ conversion factor, \xco, of
non-star-forming GMCs is also $\sim50$ per cent larger than for
star-forming GMCs. We compare the properties of GMCs with estimates
for local interstellar conditions: specifically, we investigate the
\hi\ column density, radiation field, stellar mass surface density and
the external pressure. Very few cloud properties demonstrate a clear
dependence on the environment; the exceptions are significant positive
correlations between i) the \hi\ column density and the GMC velocity
dispersion, ii) the stellar mass surface density and the average peak
CO brightness, and iii) the stellar mass surface density and the CO
surface brightness. The molecular mass surface density of GMCs without
signs of massive star formation shows no dependence on the local
radiation field, which is inconsistent with the
photoionization-regulated star formation theory proposed by
\citet{mckee89}. We find some evidence that the mass surface density
of the MAGMA clouds increases with the interstellar pressure, as
proposed by \citet{elmegreen89}, but the detailed predictions of this
model are not fulfilled once estimates for the local radiation field,
metallicity and GMC envelope mass are taken into account.
\end{abstract}


\section{Introduction}
\label{sect:intro}

\noindent In the Milky Way, molecular gas is mostly located in giant
molecular clouds (GMCs) with masses $M>10^{5}$~\msol\ \citep[][
  henceforth S87]{solomonetal87}. Understanding the physical
properties of the GMCs is important because these clouds are the
primary sites of star formation: the formation of GMCs and the
transformation of molecular gas into stars are key processes in the
life cycle of galaxies. Models of galactic evolution typically assume
that GMCs are sufficiently similar across different galactic
environments that a galaxy's star formation rate can be parameterised
as the product of the GMC formation rate and the star formation
efficiency of molecular gas
\citep[e.g.][]{ballesterosparedeshartmann07,blitzrosolowsky06}. This
approach was initially justified by studies of Galactic molecular
clouds, which found that the basic physical properties of GMCs in the
Milky Way's disc obeyed well-defined scaling relations, often referred
to as ``Larson's laws''
\citep[e.g. S87,][]{larson81,heyeretal01}. More recently, considerable
effort has been devoted to determining whether GMCs in other galaxies
also follow the Larson relations
\citep[e.g.][]{rosolowskyetal03,rosolowskyblitz05,rosolowsky07}, since
empirical evidence that GMC properties are uniform -- or at least
exhibit well-behaved correlations with a parameter such as metallicity
or pressure -- would provide valuable information for developing
models of star formation and galaxy evolution through cosmic time. \\

\noindent As well as furnishing galaxy evolution models with empirical
inputs, studies of extragalactic GMC populations aspire to resolve
long-standing questions about the physical processes that are
important for the formation and evolution of molecular clouds: are
GMCs quasi-equilibrium structures, for example, or transient features
in the turbulent interstellar medium? Do all GMCs form stars, and if
not, why not? What is the physical origin of Larson's scaling
relations?  Although a number of different theories to explain
molecular cloud properties and the Larson relations have been proposed
\citep[e.g. M89, E89, ][]{chieze87,fleck88}, there are few
extragalactic GMC samples that are comparable to the S87 catalogue,
which contains 273 clouds in the Galactic disc between longitudes
8$^{\circ}$ and 90$^{\circ}$, and with radial velocities between -100
and 200~\kms. The survey of \aco\ emission in the LMC by NANTEN
(henceforth ``the NANTEN survey'') provided the first complete
inventory of GMCs in any galaxy \citep{fukuietal08}, but did not
resolve molecular cloud structures smaller than $\sim$40~pc \citep[we
  adopt 50.1~kpc for the distance to the LMC,
  e.g.][]{alves04}. Thorough testing of the different molecular cloud
models will require deep, unbiased wide-field surveys of molecular
clouds at high angular resolution across a range of interstellar
conditions. Extensive surveys of this kind are only just feasible with
current instrumentation, and hence the number of molecular cloud
samples that can be used to falsify molecular cloud models remains
frustratingly small. \\

\noindent To date, studies of the CO emission in nearby galaxies have
concluded that extragalactic GMCs are alike. For a sample of $\sim70$
resolved GMCs located in five galaxies (M31, M33, IC10, and the
Magellanic Clouds), \citet{blitzetal07} found that extragalactic GMCs
not only follow the Galactic Larson relations, but also that different
galaxies have similar GMC mass distributions. Similar conclusions were
reached by \citet[][henceforth B08]{bolattoetal08} using a sample of
$\sim100$ resolved GMCs in twelve galaxies, although these authors
noted that molecular clouds in the SMC tend to have low CO
luminosities and narrow linewidths compared to GMCs of a similar size
in other galaxies. By comparing tracers of star formation and neutral
gas on $\sim~1$~kpc scales for galaxies in The \hi\ Nearby Galaxy
Survey \citep[THINGS, which does not include the Magellanic
  Clouds,][]{walteretal08}, \citet{leroyetal08} found that the
star-forming efficiency of molecular gas (defined as the star
formation rate surface density per unit molecular gas surface density
$SFE_{\rm H_{2}} \equiv \Sigma_{\rm SFR}/\Sigma_{\rm H_{2}}$) is
approximately constant in normal spiral galaxies, $SFE_{\rm H_{2}} =
5.25\pm2.5 \times 10^{-10}$~yr$^{-1}$. As noted by the authors, this
result could arise if the star-forming efficiency of an individual GMC
is determined by its intrinsic properties, and if the properties of
GMCs are independent of their interstellar environment
\citep[e.g.][]{krumholzmckee05}. While the existing observational
evidence has so far been interpreted in favour of uniform GMC
properties, a dependence of GMC properties on the local interstellar
environment is by no means ruled out. A constant $SFE_{\rm H_{2}}$ on
kiloparsec scales indicates that the properties of GMC ensembles are
alike on those scales; whether this conclusion can be applied to
individual GMCs is far less certain. Neither B08 nor
\citet{blitzetal07} pursued the origin of the scatter in the
extragalactic Larson relations that they observed, moreover, even
though the mean GMC mass surface density for the galaxies in their
respective samples varies by more than an order of magnitude, and the
mass surface densities of the individual GMCs varies between $\sim10$
and 1000~\mpcsq\ \citep[see also][ for evidence that the mass surface
  density of Milky Way clouds is not constant]{heyeretal09}.  A
resolved survey of a large number ($>100$) of GMCs located in a single
nearby galaxy therefore remains valuable, since it eliminates the
uncertainties inherent in combining heterogeneous datasets and
provides a sample that is large enough to investigate both the average
properties and scaling relations of an extragalactic GMC population,
as well as the dispersion around overall trends and average
quantities.\\

\noindent In this paper, we report on some initial results from the
Magellanic Mopra Assessment (MAGMA), an ongoing, high-resolution
survey of the \aco\ emission from molecular clouds in the Magellanic
Clouds using the Mopra Telescope. Here we present results from the
Large Magellanic Cloud (LMC) only; a description of the molecular
clouds surveyed by MAGMA in the Small Magellanic Cloud (SMC) has been
presented elsewhere (Muller \ea, accepted). While the \aco\ emission
from molecular gas in the LMC has been the target of extensive mapping
with the NANTEN telescope and Swedish-ESO Submillimetre Telescope
(SEST), neither project obtained observations that were ideal for
studying the Larson relations in the LMC
\citep{fukuietal08,israeletal03}. The spatial resolution of the NANTEN
survey is comparable to the size of a typical Milky Way GMC
\citep[$\sim50$~pc, e.g.][]{blitz93}; ideally, we would like to
resolve structures on smaller spatial scales in order to include less
massive GMCs in our analysis. Resolved observations are crucial,
moreover, for accurate estimates of derived GMC quantities such as
virial mass and mass surface density. The SEST Key Programme {\it CO
  in the Magellanic Clouds} mapped molecular clouds in the LMC with
comparable spatial resolution as MAGMA ($\sim10$~pc), but was strongly
biased towards regions associated with well-known sites of active star
formation. An analysis of the striking molecular cloud complex
situated south of the 30 Doradus star-forming complex using the MAGMA
data has already been presented by \citet{ottetal08} and
\citet{pinedaetal09}; in this paper, we turn our attention to the
general LMC cloud population.\\

\noindent This paper is structured as follows: in
Section~\ref{sect:data}, we summarise the MAGMA observing strategy and
our data reduction procedure, and also describe the ancillary data
that we have used in our analysis. Section~\ref{sect:cprops} outlines
the approach that we have used to identify GMCs and to measure their
physical properties. In Section~\ref{sect:younggmcs}, we compare the
properties of GMCs with and without star formation. Scaling relations
between the cloud properties are discussed in
Section~\ref{sect:larsonlaws}, while Section~\ref{sect:tracers}
presents a comparison between the intrinsic physical properties of the
GMCs and properties of the local interstellar environment. In
Section~\ref{sect:discussion}, we discuss whether our results are
consistent with i) the photoionization-regulated theory of star
formation proposed by \citet[][ henceforth M89]{mckee89} and ii) a
dominant role for interstellar gas pressure in the determination of
molecular cloud properties, as suggested by \citet[][ henceforth
  E89]{elmegreen89}. We conclude with a summary of our key results in
Section~\ref{sect:conclusions}.

\section{Data}
\label{sect:data}

\subsection{The MAGMA LMC survey: observations and data reduction}
\label{sect:observations}

\begin{figure*}
\begin{center}
\includegraphics[width=120mm,angle=270]{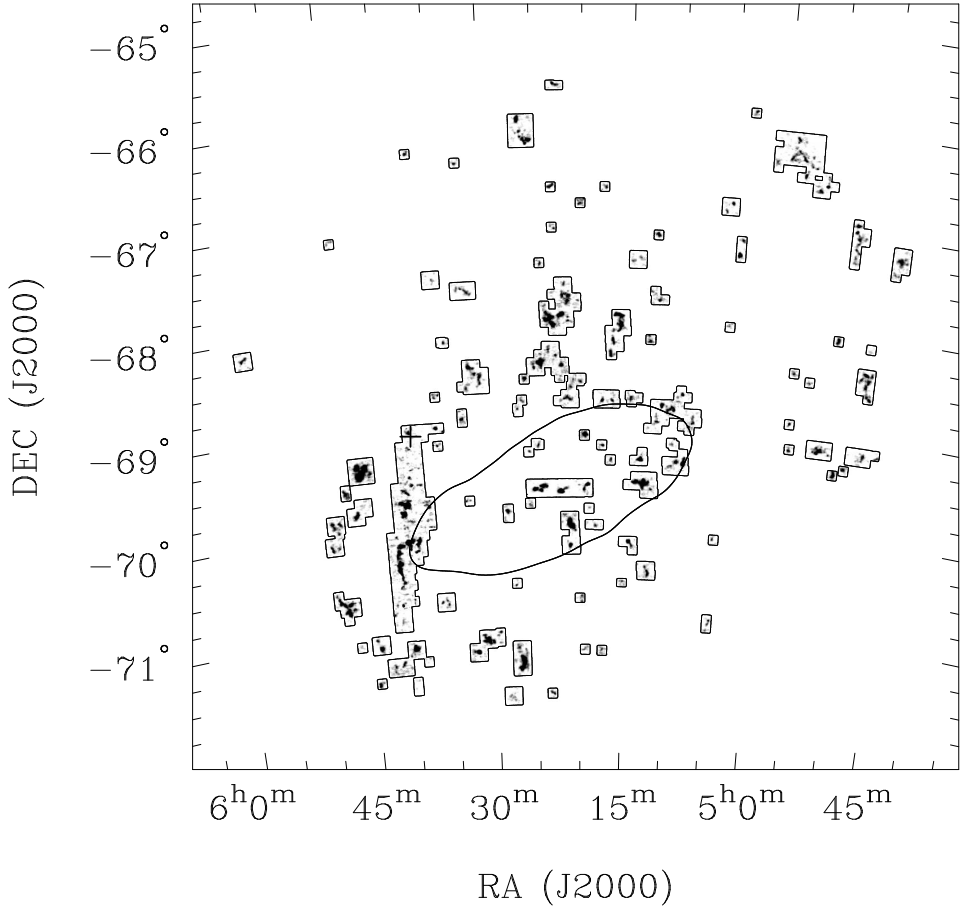}
\caption{A map of CO integrated intensity in the LMC by MAGMA
  (greyscale). The thin black lines indicate the coverage of the MAGMA
  observations included in this paper. The black cross indicates the
  position of 30~Doradus; in the text, we refer to the large survey
  region that encloses and extends south from 30~Dor as the
  ``molecular ridge''. The black ellipse indicates where the stellar
  surface density, $\Sigma_{*}$, is greater than 100~\mpcsq; we refer
  to this region in the text as the ``stellar bar''.}
\label{fig:map}
\end{center}
\end{figure*}

\noindent MAGMA observations of the \aco\ emission from molecular
clouds in the LMC are conducted at the Mopra Telescope, which is
situated near Coonabarabran, Australia.\footnote{The Mopra Telescope
  is managed by the Australia Telescope, which is funded by the
  Commonwealth of Australia for operation as a National Facility by
  the CSIRO.} At 115~GHz, the Mopra Telescope has a full width at
half-maximum (FWHM) beam size of 33 arcsec, corresponding to a spatial
resolution of 8~pc at our assumed distance to the LMC.  Due to the
large angular size of the LMC's gas disc, and the small covering
fraction of the \aco\ emission \citep[e.g.][]{mizunoetal01}, we use
the NANTEN survey of \citet{fukuietal08} to select regions of bright
\aco\ emission for high-resolution mapping. MAGMA observations target
all molecular clouds with CO luminosities $L_{\rm CO}
>7000$~\kkms~pc$^{2}$ and peak integrated intensities $>1$~\kkms\ in
the NANTEN cloud catalogue of \citet{fukuietal08}. In combination,
these clouds contribute $\sim70$ per cent of the LMC's total CO
luminosity. An integrated intensity map of the CO emission in the LMC
by MAGMA is shown in Fig.~\ref{fig:map}, with outlines that
represent the survey's coverage. \\

\noindent MAGMA observations are conducted in "on-the-fly" (OTF)
raster-mapping mode. A grid of 5 $\times$ 5 arcmin$^{2}$ OTF fields is
placed over the region surrounding a molecular cloud target. The field
centres are separated by 4.75 arcmin in RA and Dec; this 15 arcsec
overlap ensures full coverage of the molecular cloud and facilitates
mosaicking. In OTF mode, the telescope takes data continuously while
scanning across the sky. Along each row of an OTF field, individual
spectra are recorded every 14 arcsec, so that the telescope beam is
oversampled in the scanning direction. The spacing between rows is 10
arcsec, also oversampling the beam. Each row is preceded by an
off-source (emission-free) integration situated approximately 10 to 20
arcmin from the field centre; an absolute location for the OFF spectra
was selected for each molecular cloud target in order to ensure a
consistent sky subtraction. To minimize scanning artefacts, each field
in the final survey dataset is mapped twice: an initial pass is made
with the telescope scanning in the RA direction, and a second pass
then conducted in the orthogonal direction. The data presented in this
paper were obtained during the southern hemisphere winters of 2005 to
2008. Observations are complete for $\sim40$ per cent of the clouds
presented here; the remaining clouds have been scanned in the RA
direction only. The full MAGMA LMC dataset will be presented in a
subsequent paper (Wong \ea, in preparation).\\

\noindent A system temperature measurement is obtained with an ambient
load at the start of each OTF map and every 20 to 30 minutes
thereafter. In the interval between these measurements, the system
temperature is monitored using a noise diode. Typical system
temperatures for the survey observations are between 500 and 600~K;
observing is abandoned when system temperatures exceed
$\sim$1000~K. Our intention is to obtain data with uniform
sensitivity, so OTF fields with average system temperatures above
$\sim850$~K are re-observed. Between each OTF map, the pointing of the
antenna is verified by observing the nearby SiO maser RDor. The
pointing solution is updated (and re-verified) if the pointing error
in azimuth or elevation is greater than 8 arcsec. Prior to correction,
the pointing errors are typically below 10 arcsec.\\

\noindent In 2005, the Mopra Telescope was equipped with a dual
polarisation SIS receiver that produced 600~MHz instantaneous
bandwidth for observing frequencies between 86 and 115~GHz
\citep{mooreyetal97}. The correlator at that time could be configured
for bandwidths between 4 and 256~MHz across the receiver's 600~MHz
band \citep{wilsonetal92}. Our 2005 observations targeted the
molecular ridge region discussed by \citet{ottetal08} and
\citet{pinedaetal09} (see Fig.~\ref{fig:map}), with some data obtained
for additional fields near RA 05h16m, Dec -68d10m (J2000) and RA
05h24m, Dec -69d40m (J2000) (J2000). For all 2005 observations, the
correlator was configured with 1024~channels over a 64~MHz bandwidth
centred on 115.16~MHz, which provided a velocity resolution of
0.16~\kms\ per channel across a reliable velocity bandwidth of
$\sim120$~\kms. As this is not quite sufficient to cover the total
radial velocity range of the LMC's CO emission, the centre of the
observing band was placed near the peak of CO spectrum obtained by
NANTEN for the region being observed. In subsequent years, the data
were recorded using the newly installed MMIC receiver and the
University of New South Wales Digital Filter Bank (MOPS).\footnote{The
  University of New South Wales Digital Filter Bank used for the
  observations with the Mopra Telescope was provided with support from
  the Australian Research Council.} In the narrowband configuration
used for the MAGMA survey observations, MOPS can simultaneously record
dual polarisation data for up to sixteen 138~MHz windows situated
within an 8~GHz band. Each 138~MHz window is divided into 4096
channels; for our survey observations at 115~GHz, this configuration
provides a velocity resolution of 0.09~\kms\ per channel across the
velocity range [90,410]~\kms. \\

\noindent The Mopra beam has been described by
\citet{laddetal05}. These authors identify three components - the
'main beam', and the inner and outer 'error beams' - that contribute
to the antenna response. The presence of these non-negligible error
beams implies that the telescope efficiency, $\eta$, will depend on
source size. In the 2004 observing season, \citet{laddetal05} derived
a main-beam efficiency factor of $\eta_{\rm mb} = 0.42$ for
observations at 115~GHz, and an 'extended' beam efficiency of
$\eta_{\rm xb} = 0.55$ for sources that are comparable to the size of
the inner error beam ($\sim80$ arcsec). For clouds in the MAGMA
survey, we consider the extended beam efficiency factor to be more
appropriate.\\

\noindent To monitor the reliability of the flux calibration, we
observed the standard source Orion~KL (RA 05h35m14.5s, Dec
-05d22m29.56s (J2000)) once per observing session throughout our
survey. In 2005, we measured an average peak brightness temperature of
Orion~KL of $55\pm3$~K (in $T_{A}^{*}$ units). According to the SEST
documentation,\footnote{{\tt
    http://www.apex-telescope.org/sest/html/telescope-calibration/calib-sources/orionkl.html}}
the peak antenna temperature of Orion~KL is $T_{A}^{*}\sim71$~K,
corresponding to a main beam temperature of $T_{\rm mb} \sim 102$~K
for an efficiency factor of $\eta_{\rm mb}=0.7$
\citep{johanssonetal98}. This suggests that data from the two
telescopes can be placed on the same brightness temperature scale
using a factor of $\eta_{\rm xb}=0.54\pm0.03$ for the Mopra data, in
excellent agreement with the telescope efficiency derived by
\citet{laddetal05}.  In subsequent years, the average peak brightness
temperature that we have measured for Orion~KL has varied. In 2006, we
measured an average peak brightness temperature of $35\pm4$~K, while
our average measurements in 2007 and 2008 were $43\pm3$~K and
$50\pm3$~K respectively. For these data, we therefore used conversion
factors of $0.35$ (2006), $0.43$ (2007) and $0.49$ (2008). To validate
the final flux scale of our Mopra data, we compiled published SEST
measurements of LMC molecular clouds from the studies by
\citet{israeletal93}, \citet{chinetal97}, \citet{kutneretal97},
\citet{johanssonetal98} and \citet{israeletal03}. We were able to
compare Mopra and SEST measurements of the peak brightness and peak
integrated intensity for 40 clouds. For both the peak brightness and
the peak integrated intensity of these clouds, the average ratio
between the SEST and Mopra measurements was $1.1$, with a dispersion
of $\sim20$ per cent. \\

\begin{figure}
\begin{center}
\includegraphics[width=70mm,angle=270]{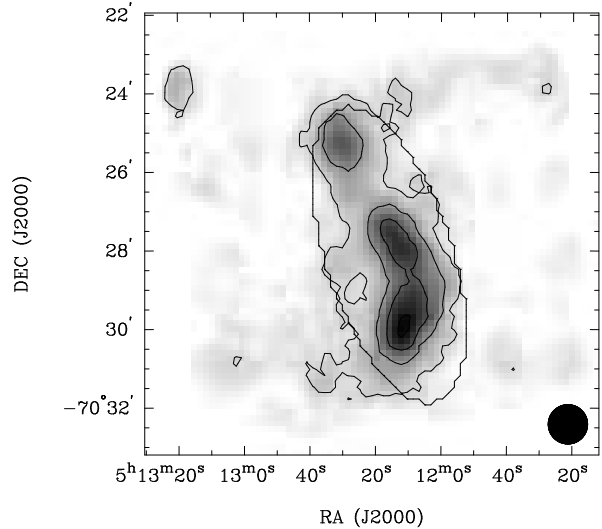}
\caption{A \aco\ integrated intensity map of an example GMC in the
  MAGMA cloud list. The contour spacing is 2~\kms, with the lowest
  contour at $I_{\rm CO} = 1.5$~\kms. The ellipse represents the fit
  derived by \textsc{CPROPS}, from which the major and minor axis and
  position angle of the GMC were taken. As explained in
  Section~\ref{sect:cprops}, we are interested to identify GMCs in the
  MAGMA data subcubes, rather than structure on smaller scales
  corresponding to the spatial resolution of the survey. The black
  circle in the lower right corner represents the Mopra beam.}
\label{fig:egmom}
\end{center}
\end{figure}

\noindent The processing of the survey data involved four main
steps. An initial correction to align the position and time stamp
information in the raw data files was applied to data obtained in
2005, using the {\tt mopfix} task of the \textsc{MIRIAD} software
package \citep{saultetal95}. Bandpass calibration and baseline-fitting
were performed using the \textsc{AIPS++ livedata}
package. \textsc{Livedata} determines the bandpass calibration for
each row of an OTF map using the preceding OFF scan and then fits a
user-specified polynomial to the spectral baseline. We chose to fit
the spectral baselines with a first-order polynomial: data subcubes
containing baselines showing higher order ripples were rejected. The
spectra were combined to form a spectral line cube using the
\textsc{AIPS++ gridzilla} package. Each $(x,y,v)$ cell within the
final cube is sampled multiple times by the OTF scans:
\textsc{gridzilla} averages the spectra contributing to the emission
within each cell according to a convolution kernel, beam profile and
weighting scheme specified by the user. We chose to grid the data with
a cell size of 9 $\times$ 9 arcsec$^{2}$, aggregating the data from
both polarisations and both scanning directions. The data were
weighted by the inverse of the system temperature measurements. We
used a truncated Gaussian convolution kernel with a FWHM of 1 arcmin
and a cutoff radius of 30 arcsec, producing an output data subcube
with effective angular resolution of 45 arcsec. Although
\textsc{gridzilla} can process data from contiguous OTF fields, it
cannot process data recorded with different correlator configurations
simultaneously. The data subcubes from different years were thus
processed separately, and were converted to $T_{\rm mb}$ units using
the annual extended beam efficiency factors derived from our Orion~KL
observations. The \textsc{MIRIAD} task {\tt imcomb}, which weights the
input data by the inverse of the RMS noise, was then used to combine
data from regions that were observed over multiple observing
seasons. Finally, in order to increase the signal-to-noise ($S/N$), we
binned the MAGMA data subcubes to a velocity resolution of
0.53~\kms\ and smoothed them to an angular resolution of 1 arcmin. The
average RMS noise in these subcubes is 0.24~K per 0.53~\kms
channel. An example of the integrated \aco\ emission for one such
MAGMA subcube is shown in Fig.~\ref{fig:egmom}.\\

\subsection{Ancillary data}
\label{sect:ancillary}

\subsubsection{\hi\ data}
\label{sect:himap}

\noindent To trace the atomic gas in the LMC, we use the \hi\ map
published by \citet{kimetal03}, which combines data from the Australia
Telescope Compact Array \citep[ATCA][]{kimetal98} and the Parkes
single-dish telescope \citep{staveleysmithetal03}. The angular
resolution of the \hi\ data is 1 arcmin, well-matched to the reduced
MAGMA CO data. The \hi\ data cube has a velocity resolution of
1.65~\kms, with a column density sensitivity of $7.2 \times
10^{18}$~\pcmsq\ per channel. We construct an \hi\ integrated
intensity emission map of the LMC, $I_{\hi}$, by integrating the
\hi\ data cube over the heliocentric velocity range 196 to
353~\kms. We assume that the \hi\ emission is optically thin
everywhere, and derive a map of the LMC's \hi\ column density, $\nh$,
according to
\begin{equation}
\nh [\pcmsq] = 1.82 \times 10^{18} I_{\hi} [\kkms].
\end{equation}
\noindent Our estimate for \nh\ is likely to be a lower limit
since the \hi\ emission in the LMC may have significant optical depth,
especially along the sightlines towards molecular clouds \citep[][ see
  also
  Section~\ref{sect:e89}]{dickeyetal94,marxzimmeretal00,bernardetal08trunc}.

\subsubsection{Stellar mass surface density}
\label{sect:starmap}

\noindent To trace the mass distribution within the LMC's stellar
disc, we use the stellar mass surface density map presented in
fig.~1c of \citet{yangetal07}. The map is based on number counts of
red giant branch (RGB) and asymptotic giant branch (AGB) stars,
selected by their colours from the Two Micron All Sky Survey Point
Source Catalogue \citep{skrutskieetal06trunc}. The star counts are binned
into 40~pc $\times$ 40~pc pixels, and then convolved with a Gaussian
smoothing kernel with $\sigma = 100$~pc. The resulting map is
normalized to a measure of absolute stellar mass surface density by
adopting a total stellar mass for the LMC of $2 \times
10^{9}$~\msol\ \citep{kimetal98}. The resolution of the stellar
surface density map is considerably coarser than the angular
resolution of our MAGMA data subcubes, so it is possible that the
average stellar mass surface density of smaller GMCs is underestimated
due to beam dilution. RGB and AGB stars are relatively old
populations, however, so their spatial distribution is likely to be
smooth. In particular, we do not expect them to be strongly clustered
in the vicinity of molecular clouds
\citep[e.g.][]{nikolaevweinberg00}.

\subsubsection{Interstellar radiation field}
\label{sect:isrfmap}

\noindent To estimate the interstellar radiation field at the
locations of molecular clouds within the LMC, we use the dust
temperature ($T_{d}$) map presented in fig.~7 of
\citet[][]{bernardetal08trunc}. The map is derived using the ratio of the
IRIS 100~$\mu$m and {\it Spitzer} 160~$\mu$m emission maps
\citep{mivilledescheneslagache05,meixneretal06trunc}, assuming that dust
emission can be modelled as a grey body:
\begin{equation}
I_{\nu} \propto \nu^{\beta}B_{\nu}(T_{d}), 
\end{equation}
\noindent with $\beta=2$ at far-infrared wavelengths.\footnote{'IRIS'
  is an abbreviation for Improved Reprocessing of the {\it IRAS}
  Survey. As described in \citet{mivilledescheneslagache05}, IRIS
  images have improved zodiacal light subtraction, absolute
  calibration and scanning stripe suppression than the {\it IRAS} Sky
  Survey images.} \citet{bernardetal08trunc} find that dust
temperatures in the LMC vary from 12~K up to 34.7~K, with an average
value of 18.3~K. This is significantly colder than previous
determinations, especially for estimates that attempted to constrain
the dust temperature using the {\it IRAS} 60$\mu$m flux density. As
noted by \citet{bernardetal08trunc}, emission at 60$\mu$m is highly
contaminated by out-of-equilibrium emission from very small grains
(VSGs), and this is especially true in the LMC, due to the presence of
excess 70$\mu$m emission. Temperatures derived from the {\it IRAS}
60/100$\mu$m flux density ratio may therefore be strongly
overestimated.\\

\noindent For dust grains in thermal equilibrium, the strength of the
radiation field, $G_{0}$, is related to the dust temperature by $G_{0}
\propto T_{d}^{4+\beta}$ \citep[e.g.][]{lequeuxbook}. The average
strength of $G_{0}$ across the entire LMC is thus only a factor of
$\sim1.3$ greater than in the solar neighbourhood, $G_{0,\odot}$,
where the dust temperature is 17.5~K \citep[with
  $\beta=2$,][]{boulangeretal96}. At the locations observed by MAGMA,
$G_{0}/G_{0,\odot}$ varies between 0.5 and 58.8, with a median value
of 1.7. For $G_{0}/G_{0,\odot}$ averaged over the projected cloud
areas, we obtain values in the range 0.5 to 8.0. Our derived value for
the strength of the radiation field near the well-known LMC
star-forming region N159W ($G_{0}/G_{0,\odot}$ = 8.4) is very low
compared to the value obtained by previous estimates
\citep[e.g. ][]{israeletal96,pinedaetal08trunc,pinedaetal09}. Part of
the discrepancy may be due to beam dilution since the resolution of
the \citet{bernardetal08trunc} dust temperature map is 4 arcmin, which
corresponds to a spatial scale of 60~pc in the LMC. However, we also
note that our method for determining $G_{0}/G_{0,\odot}$ is extremely
sensitive to the assumed dust temperature, which is colder than
reported by previous analyses. At the locations of the molecular
clouds in the \citet{bernardetal08trunc} dust temperature map, the
mean (maximum) formal error on $T_{d}$ is 2 per cent (12 per
cent). These errors do not include potential variations in the value
of $\beta$, however, so the true uncertainty is likely to be greater
than this. Despite the uncertainties regarding the absolute value of
$G_{0}/G_{0,\odot}$, we emphasise that our analysis concerns the
relationship between GMC properties and the relative strength of the
radiation field, and our conclusions do not rely on an absolute
calibration of $G_{0}/G_{0,\odot}$. Finally, we note that while the
resolution of the \citet{bernardetal08trunc} dust temperature map is
coarser than the angular resolution of our MAGMA CO data, it is still
likely to provide a reasonable estimate for the average dust
temperature within individual GMCs with $R\geq 30$~pc. For smaller
clouds, the average dust temperature and average radiation field may
be underestimated due to beam dilution.

\section{Measuring GMC properties}
\label{sect:cprops}

\noindent To identify GMCs in the MAGMA data subcubes and measure
their properties, we have used the algorithms presented in
\citet[][implemented in \textsc{IDL} as part of the \textsc{CPROPS}
  package]{rosolowskyleroy06}. \textsc{CPROPS} uses a dilated mask
technique to isolate regions of significant emission within spectral
line cubes, and a modified watershed algorithm to assign the emission
into individual clouds. Moments of the emission along the spatial and
spectral axes are used to determine the size, linewidth and flux of
the clouds, and optional corrections for the finite sensitivity and
instrumental resolution may be applied to the measured cloud
properties. Each step of the \textsc{CPROPS} method is described in
detail by \citet{rosolowskyleroy06}. \\

\begin{figure*}
\begin{center}
\includegraphics[width=100mm,angle=270]{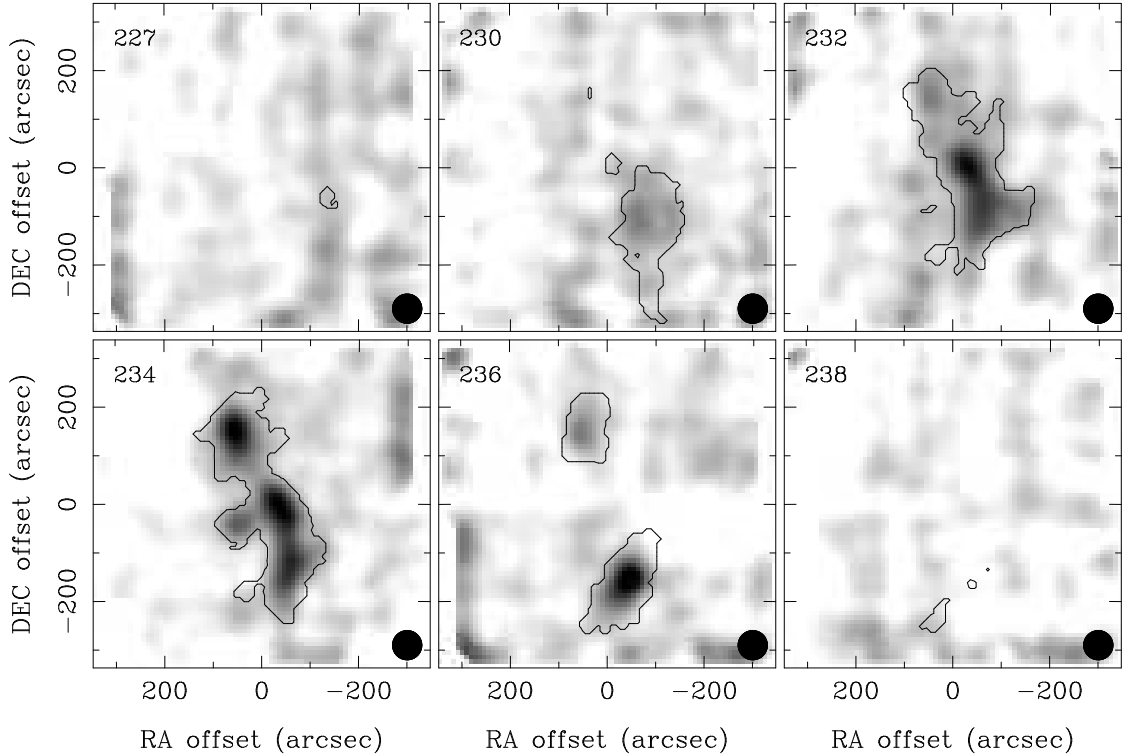}
\vspace{0.5cm}
\caption{Channel maps of the CO emission from the GMC in
  Fig.~\ref{fig:egmom} (greyscale). The black contour indicates the
  emission region that \textsc{CPROPS} identifies as belonging to the
  cloud. The velocity axis of the MAGMA data subcube and
  \textsc{CPROPS} assignment cube has been binned to a channel width
  of 2.1~\kms\ for illustration only. The black circle in the lower
  right corner of each panel represents the Mopra beam.}
\label{fig:egchan}
\end{center}
\end{figure*}

\noindent Regions of significant emission within the MAGMA data subcubes
are initially identified by finding pixels with emission greater than
a threshold of $4\sigma_{RMS}$ across two contiguous velocity
channels. The mask around these core regions is then expanded to
include all the pixels connected to the core with emission greater
$1.5\sigma_{RMS}$ across at least two consecutive channels. We
explored a range of values for the threshold and edge parameters in
the masking process, and found that these values distinguished
credible emission regions (i.e. the mask did not expand excessively
into the noise) and also yielded reliable measurements for the
properties of faint clouds. The emission identified with an isolated
cloud in the MAGMA dataset is illustrated in Fig.~\ref{fig:egchan}.  \\

\noindent Once regions of significant emission have been identified,
\textsc{CPROPS} assigns the emission to individual cloud
structures. To generate the preliminary list of GMC candidates, we
used the default parameters for the identification of GMCs that are
recommended in \citet{rosolowskyleroy06}. In this case, the parameters
of the decomposition are motivated by the observed physical properties
of Galactic GMCs: spatial sizes greater than $\sim10$~pc, linewidths
of several \kms, and brightness temperatures less than $\sim10$~K. We
adopt this approach because our goal is to describe the properties of
GMCs in the LMC and to investigate how these properties might differ
to the properties of GMCs in other galaxies. As noted by
\citet{rosolowskyleroy06}, GMCs contain structure across a wide range
of size scales, so identifying the clumpy substructure -- with a size
scale of $\sim1$~pc and typical linewidth of $\sim1$~\kms\ -- within
the clouds in our spectral line cubes would require different
decomposition parameters than the ones that we have used. While the
properties and scaling relations of this substructure is an important
topic for investigation, we defer this analysis to a future paper. A
previous analysis of the MAGMA data in the molecular ridge region
employed a different decomposition algorithm
\citep[i.e. \textsc{GAUSSCLUMPS}, ][]{pinedaetal09}, which identified
structures that are typically smaller (with radii between 6 and 20~pc)
than the GMCs described here.\\

\noindent Our initial list of GMC candidates contained 237
objects. After cross-checking the \textsc{CPROPS} cloud
identifications against the NANTEN \aco\ data cube and the MAGMA data
subcubes to verify that the identified objects were genuine, we
rejected 54 features that were clearly noise peaks or map edge
artefacts. To ensure that the properties of clouds in our final list
are reliable, we impose a signal-to-noise threshold $S/N \ge 5$ and
reject objects with measurement errors for the radius and velocity
dispersion that are greater than 20 per cent. We further require that
\textsc{CPROPS} is able to apply the corrections for sensitivity and
instrumental resolution successfully (i.e. these corrections do not
result in undefined values). The resulting sample of GMCs, which we
refer to as the ``MAGMA cloud list'', contains 125 clouds. To verify
the results of our analysis, we also define a ``high quality'' GMC
sample, which contains 57 clouds with $S/N \ge 9$ and for which the
uncertainties in the radius and velocity dispersion measurements are
less than 15 per cent. \\

\noindent Cloud properties follow the standard \textsc{CPROPS}
definitions. The cloud radius is defined as $R = 1.91 \sigma_{R}$,
where $\sigma_{R}$ is the geometric mean of the second moments of the
emission along the cloud's major and minor axes. The velocity
dispersion, $\sigma_{\rm v}$, is the second moment of the emission
distribution along the velocity axis, which for a Gaussian line
profile is related to the FWHM linewidth, $\Delta v$, by $\Delta v =
\sqrt{8 \ln 2} \ts\sigma_{\rm v}$ \citep[e.g.][]{rosolowskyleroy06}. The
CO luminosity of the cloud, $L_{\rm CO}$, is simply the emission
inside the cloud integrated over position and velocity, i.e.
\begin{equation} 
L_{\rm CO} ({\rm K~\kms~pc}^{2}) = D^{2} \left( \frac{\pi}{180 \times 3600}\right)^{2} \Sigma T_{i}
\delta v \delta x \delta y
\end{equation}
\noindent where $D$ is the distance to the LMC in parsecs, $\delta x$
and $\delta y$ are the spatial dimensions of a pixel in arcseconds,
and $\delta v$ is the width of one channel in \kms. The virial mass is
given by $M_{\rm vir} = 1040 \sigma_{\rm v}^{2}R$~\msol, which assumes
that molecular clouds are spherical with $\rho \propto r^{-1}$ density
profiles \citep{maclarenetal88}. \textsc{CPROPS} estimates the error
associated with a cloud property measurement using a bootstrapping
method, which is described in section~2.5 of
\citet{rosolowskyleroy06}.  \\

\noindent In addition to the basic properties reported by
\textsc{CPROPS}, we define the CO surface brightess as the total CO
luminosity of a cloud divided by its area, $I_{\rm CO} [\kkms] \equiv
L_{\rm CO}/\pi R^{2}$. The molecular mass surface density,
$\Sigma_{\rm H_{2}}$, is defined as the virial mass divided by the
cloud area, $\Sigma_{\rm H_{2}} [\mpcsq] \equiv M_{\rm vir}/\pi
R^{2}$. The CO-to-\hh\ conversion factor, \xco, may be expressed as
the ratio between these quantities: $\xco\ [\xcou] \equiv \Sigma_{\rm
  H_{2}} [\mpcsq] / (2.2 I_{\rm CO} [\kkms])$. We note that our
definitions of \xco\ and $\Sigma_{\rm H_{2}}$ assume that GMCs manage
to achieve dynamic equilibrium: if the degree of virialisation is not
constant for LMC molecular clouds, then variations in \xco\ and
$\Sigma_{\rm H_{2}}$ may instead reflect differences between the
dynamical state of the clouds. We estimate the uncertainties in \xco,
$I_{\rm CO}$ and $\Sigma_{\rm H_{2}}$ using standard error propagation
rules.\\

\noindent As emphasized by \citet{rosolowskyleroy06}, the resolution
and sensitivity of a dataset influence the derived cloud
properties. In order to reduce these observational biases, they
recommend extrapolating the cloud property measurements to values that
would be expected in the limiting case of perfect sensitivity (i.e. a
brightness temperature threshold of 0~K), and correcting for finite
resolution in the spatial and spectral domains by deconvolving the
telescope beam from the measured cloud size, and deconvolving the
channel width from the measured linewidth. The procedures that
\textsc{CPROPS} uses to apply these corrections are described in
\citet[][see especially fig.~2 and sections 2.2 and
  2.3]{rosolowskyleroy06}. For the analysis in this paper, we use
cloud property measurements that have been corrected for resolution
and sensitivity bias. \\

\noindent The GMCs observed by MAGMA in the LMC have radii ranging
between 13 and 160~pc, velocity dispersions between 1.0 and 6.1~\kms,
peak CO brightnesses between 1.2 and 7.1~K, CO luminosities between
$10^{3.5}$ and $10^{5.5}$~K~\kms~pc$^{2}$, and virial masses between
$10^{4.2}$ and $10^{6.8}$~\msol. The clouds tend to be elongated, with
a median axis ratio of 1.7. A detailed analysis of the cloud property
distributions and their uncertainties will be presented once the MAGMA
LMC survey is complete (Wong \ea, in preparation). 

\section{Physical properties of non-star-forming GMCs}
\label{sect:younggmcs}

\noindent Molecular clouds have traditionally been modelled as
quasi-equilibrium structures, but recent theoretical work has also
begun to explore whether molecular clouds might form and disperse more
rapidly as a consequence of large-scale dynamical events in the ISM,
such as turbulent flows or cloud collisions
\citep[e.g.][]{berginetal04,taskertan08}. Observational constraints on
the physical properties of recently formed GMCs would be a useful
contribution to the debate about GMC lifetimes, but it remains unclear
whether observations of \aco\ emission alone can distinguish between
younger and more evolved GMCs. While it is plausible that some
potential characteristics of newly formed GMCs, such as colder gas
temperatures, stronger bulk motions or sparser CO filling factors,
would have observational signatures, it is also possible that the
onset of widespread \aco\ emission might occur late in the cloud
formation process, or that physical conditions in the CO-emitting
regions of GMCs are relatively uniform and therefore insensitive to a
cloud's evolutionary state. In this section, we investigate whether
there are significant differences between the properties of young GMC
candidates and other GMCs in the MAGMA LMC cloud list. \\

\noindent We constructed a sample of young GMC candidates using the
evolutionary classification scheme designed by \citet{kawamuraetal09}
for the 272 GMCs in the NANTEN LMC catalogue
\citep{fukuietal08}. \citet{kawamuraetal09} classified GMCs on the
basis of their association with \hii\ regions and young stellar
clusters, finding 72 Type I GMCs that show no association with massive
star-forming phenomena. An important assumption behind this approach
is that all GMCs eventually form stars and are finally dissipated by
their stellar offspring; in this scenario, GMCs without signs of
massive star formation may be considered young. To date, MAGMA has
observed 30 Type I NANTEN GMCs, but there are only 17 MAGMA clouds
associated with these Type I GMCs that satisfy our criteria for
inclusion in the MAGMA cloud list: henceforth, we refer to these 17
MAGMA clouds as the ``young GMC sample'' or the ``non-star-forming
GMCs''. The 17 clouds in the young GMC sample correspond to 16 NANTEN
GMCs: 15 of the 17 MAGMA clouds demonstrate a one-to-one
correspondence with a Type I GMC in the NANTEN catalogue, while the
other NANTEN GMC divides into two clouds at MAGMA's finer
resolution. \textsc{CPROPS} identifies 10 more objects in the MAGMA
data subcubes that are associated with a further 8 NANTEN Type I GMCs,
but these objects either have $S/N < 5$ and/or measurement errors that
exceed 20 per cent, and are therefore excluded from the MAGMA cloud
list. \textsc{CPROPS} does not identify any significant emission in
the MAGMA data subcubes for the remaining six NANTEN Type I GMCs that
we have observed to date, although we note that these are all regions
of the MAGMA survey that have only been scanned once. We do not
attempt to re-classify the MAGMA clouds according to the criteria
developed by \citet{kawamuraetal09}, but simply ascribe the
evolutionary classification of the NANTEN GMC to any MAGMA cloud that
is coincident in space and velocity. This classification should be
reliable for clouds in the young GMC sample since their projected
areas are always smaller than, and contained within, the corresponding
NANTEN GMC boundary.\\

\noindent To verify that differences between the properties of
star-forming and non-star-forming GMCs are not due solely to
variations in the cloud size, we constructed a control sample with the
same size distribution as the young GMC sample by matching each of the
17 young GMC candidates with three star-forming clouds in the MAGMA
cloud list of a similar radius (i.e. the control sample contains 51
clouds). \citet{kawamuraetal09} found that young GMCs in the NANTEN
catalogue tend to be smaller than star-forming GMCs, but here we wish
to determine whether there are differences between star-forming and
non-star-forming GMCs that persist even after variations in cloud size
are suppressed. The average discrepancy between the radius of a young
GMC candidates and their three corresponding control clouds was 2 per
cent, with a maximum discrepancy of 8 per cent. The average properties
of GMCs in the MAGMA cloud list, the young GMC sample and the control
sample are listed in Table~\ref{tbl:ks1}.\\

\begin{table*}
\begin{minipage}{170mm}
\caption{Average physical properties of the MAGMA clouds and results
  of the KS tests for the young GMC sample. Columns 2 to 4 list the
  median and median absolute deviation of th properties of clouds in
  the MAGMA cloud list, the young GMC sample and the control sample
  respectively. Column 5 lists the median $P$ value obtained in the
  error trials, and column 6 lists the standard deviation of the $P$
  values in the trials (see text).  }
\label{tbl:ks1} 								        
\begin{tabular}{@{}lccccc}
\hline
Cloud Property   & MAGMA & Young GMCs & Control & $\langle P \rangle$ & $\sigma(P)$ \\
\hline
$R$ (pc)	                        &  28$\pm$8     & 30$\pm$8    & 30$\pm$7      & 0.96  &  0.05   \\   
$\sigma_{\rm v}$ (\kms)                 &  2.3$\pm$0.5 & 2.4$\pm$0.7  & 2.1$\pm$0.5    & 0.54  &  0.18   \\    
$\langle T_{\rm pk} \rangle$ (K)        &  1.1$\pm$0.1 & 1.0$\pm$0.1  & 1.1$\pm$0.1    & 0.007  &  0.007    \\
$T_{\rm max}$ (K) 	                &  2.2$\pm$0.4 & 1.7$\pm$0.2  & 2.3$\pm$0.4   & 0.003  &  0.003 \\
$L_{\rm CO}$ ($10^{4}$~K~\kms~pc$^{2}$) &  1.2$\pm$0.6 & 1.2$\pm$0.5  & 1.2$\pm$0.5      &  0.89  &  0.11 \\
$M_{\rm vir}$ ($10^{5}$~\msol)	        &  1.5$\pm$0.9 & 1.7$\pm$1.0  & 1.5$\pm$0.8   & 0.78  &  0.18  \\
Axis Ratio, $\Gamma$	                &  1.7$\pm$0.4 & 1.5$\pm$0.2  & 1.8$\pm$0.4   & 0.25  &  0.10  \\
$\Sigma_{\rm H_{2}}$ (\msol~pc$^{-2}$)  &  55$\pm$22   & 73$\pm$21    & 52$\pm$23         & 0.13  &  0.10  \\
$I_{\rm CO}$ (K~\kms)                   &  4.8$\pm$1.5 & 4.3$\pm$0.8  & 4.6$\pm$1.3     & 0.78  &  0.15  \\
$\xco$ ($\xcou$)	                &  4.7$\pm$1.6 & 6.9$\pm$2.3  & 5.4$\pm$2.0   & 0.06  &  0.05  \\
$V_{LSR} $ (\kms)                      &  255$\pm$22  & 248$\pm$18   & 256$\pm$23      & 0.18  &  0.00	\\
$R_{\rm gal}$ (kpc)                     &  1.8$\pm$0.9 & 2.3$\pm$0.9  & 1.8$\pm$0.8     & 0.43  &  0.00	\\
\nh\ ($\times 10^{21}$~\pcmsq)      &  2.7$\pm$0.7 & 2.7$\pm$0.4  & 2.7$\pm$0.7        & 0.66  &  0.18	\\
$\Sigma_{*}$ (\mpcsq)                   &  48$\pm$24   & 50$\pm$16    & 42$\pm$19      & 0.25  &  0.10	 \\
$G_{0}/G_{0,\odot}$                     &  1.7$\pm$0.5 & 1.3$\pm$0.5  & 2.0$\pm$0.6      & 0.003 & 0.003	\\
$\langle T_{\rm pk}(\hi) \rangle$ (K)   &  68$\pm$10   & 51$\pm$10    & 68$\pm$8        & 0.0001  &  0.0001 \\
$P_{h}/k_{\rm B}$ ($10^{4}$~K~\ccc)      &  6.4$\pm$2.5   & 6.4$\pm$0.9  & 5.54$\pm$2.1  & 0.54  & 0.18 \\
\hline
\end{tabular}
\end{minipage}
\end{table*}

\noindent To test whether young GMCs have distinct physical
properties, we conducted Kolmogorov-Smirnov (KS) tests between the
young GMC sample and the control sample. The KS test is a
non-parametric test that compares the cumulative distribution
functions of two samples in order to determine whether there is a
statistically significant difference between the two populations. The
test is reliable when the effective number of data points, $N_{e}
\equiv \frac{N_{1}N_{2}}{N_{1}+N_{2}}$, is greater than 4, where
$N_{1}$ and $N_{2}$ are the number of data points in the first and
second samples. The result of the KS test can be expressed as a
probability, $P$, that the sample distributions are drawn from the
same underlying distribution. As the traditional KS test does not
account for uncertainties in the measurements of the cloud properties,
we performed 1000 trials of each KS test. For each trial, the value of
each cloud property measurement was displaced by $k\Delta x$, where
$\Delta x$ is the absolute uncertainty in the cloud property
measurement and $k$ is a uniformly distributed random number between
-1 and 1. A summary of the KS test results is shown in
Table~\ref{tbl:ks1}. If the results of the error trials are narrowly
distributed around zero -- i.e. if $\langle P \rangle \leq 0.05$ and
$\sigma(P) \leq 0.05$, where $\sigma$ represents the standard deviation
-- we consider that there is statistically significant evidence
against the null hypothesis that the cloud samples are drawn from the
same underlying population. \\

\noindent Table~\ref{tbl:ks1} shows that the properties of both the
young GMC and the control sample show a large dispersion around their
average value, and that differences between the samples are not always
obvious from measures of central tendency. The results of the KS tests
indicate that the null hypothesis cannot be rejected for many of the
intrinsic cloud properties. Both the maximum peak CO brightness within
a GMC ($T_{\rm max}$) and the average peak CO brightness ($\langle
T_{\rm pk} \rangle$, i.e. the peak CO intensity at the line centre for
all independent sightlines through the cloud, averaged over the
projected cloud area) tend to be lower for young GMCs. There is no
significant difference between the distributions of the total CO
luminosity ($L_{\rm CO}$) or CO surface brightness ($I_{\rm CO}$)
however. The median \xco\ value for the young GMC sample ($\langle
\xco \rangle = 6.9 \times 10^{20}$~\xcou) is $\sim50$ per cent greater
than the median \xco\ value for the entire MAGMA LMC cloud list. Some
of this variation may be due to cloud size, since the discrepancy is
reduced once we restrict our comparison to the control sample. The KS
test results indicate that the difference is only marginally significant.\\

\noindent For properties of the local interstellar environment, young
GMCs appear to be distributed throughout the LMC in a similar fashion
as star-forming GMCs (i.e. across a similar range of galactocentric
radii and radial velocities), and they are detected across a
comparable range of stellar mass surface densities and \hi\ column
densities. There is a clear trend, however, for young GMCs to be
located in regions where the \hi\ peak brightness ($\langle T_{\rm
  pk}(\hi) \rangle$) is relatively low, and the radiation field
($G_{0}$) is relatively weak. In Section~\ref{sect:fukuigmcmodel}, we
argue that the latter reflects the fact that young massive stars are
an important source of dust heating in LMC molecular clouds.

\section{Cloud scaling relations}
\label{sect:larsonlaws}

\noindent Empirical scaling relations between the basic physical
properties of molecular clouds have become a standard test for
potential differences between molecular cloud populations. Larson's
initial work identified the three well-known ``laws'' obeyed by
Galactic molecular clouds: i) a power-law relationship between the
size of a cloud and its velocity dispersion, ii) a nearly linear
correlation between the virial mass of a cloud and mass estimates
based on molecular line tracers of \hh\ column density, which seemed
to imply that molecular clouds are self-gravitating and in approximate
virial balance, and iii) an inverse relationship between the size of a
cloud and its average density \citep{larson79,larson81}. S87 were
subsequently able to measure the coefficients and exponents of the
power-law relationships between the properties of 273 GMCs in the
inner Milky Way, establishing the empirical expressions for Larson's
Laws that have become the yardstick for studies of GMCs in other
galaxies and in different interstellar environments \citep[see
  e.g. B08,][]{blitzetal07}. \\

\noindent Most famously, the analysis by S87 showed that inner Milky
Way clouds follow a size-linewidth relation of the form $\sigma_{\rm
  v} = 0.72R^{0.5\pm0.05}$~\kms, and that there is a strong,
approximately linear correlation between the clouds' virial mass and
their CO luminosity, $M_{\rm vir} = 39 L_{\rm
  CO}^{0.81\pm0.03}$~\msol. These two relations can be combined to
provide expressions for the relationship between a cloud's luminosity
and size, and luminosity and linewidth: $L_{\rm CO} \approx 25
R^{2.5}$~\kkms~pc$^{-2}$ and $L_{\rm CO} \approx 130 \sigma_{\rm
  v}^{5}$~K~\kms~pc$^{-2}$. As noted by S87, clouds in gravitational
equilibrium ($M \propto R\sigma_{\rm v}^{2}$) with $\sigma_{\rm v}
\propto R^{0.5}$ will follow a mass-size relation of the form $M
\propto R^{2}$. This implies constant average mass surface density for
molecular clouds, $\langle \Sigma_{\rm H_{2}} \rangle$, which is
related to the coefficient of the size-linewidth relation, $C_{0}
\equiv \sigma_{\rm v}/\sqrt{R}$, via $\langle \Sigma_{\rm H_{2}}
\rangle \approx 331 C_{0}^{2}$~\mpcsq\ for the cloud density profile
that we have adopted. S87 found $C_{0} = 0.72$~\kms~pc$^{-0.5}$ and
$\langle \Sigma_{\rm H_{2}} \rangle \sim 170$~\mpcsq\ for molecular
clouds in the inner Milky Way, slightly higher than the median surface
density of the extragalactic GMCs analysed by B08 ($\langle
\Sigma_{\rm H_{2}} \rangle \sim 130~\mpcsq$). \citet{heyeretal09} have
recently argued that the S87 estimate should be revised downwards to
$\sim100$~\mpcsq, and we adopt this as our reference value in
Figs.~\ref{fig:Rdv} to~\ref{fig:phcmp}.

\subsection{The size-linewidth relation}
\label{sect:Rdv}

\begin{figure*}
\begin{center}
\includegraphics[width=70mm]{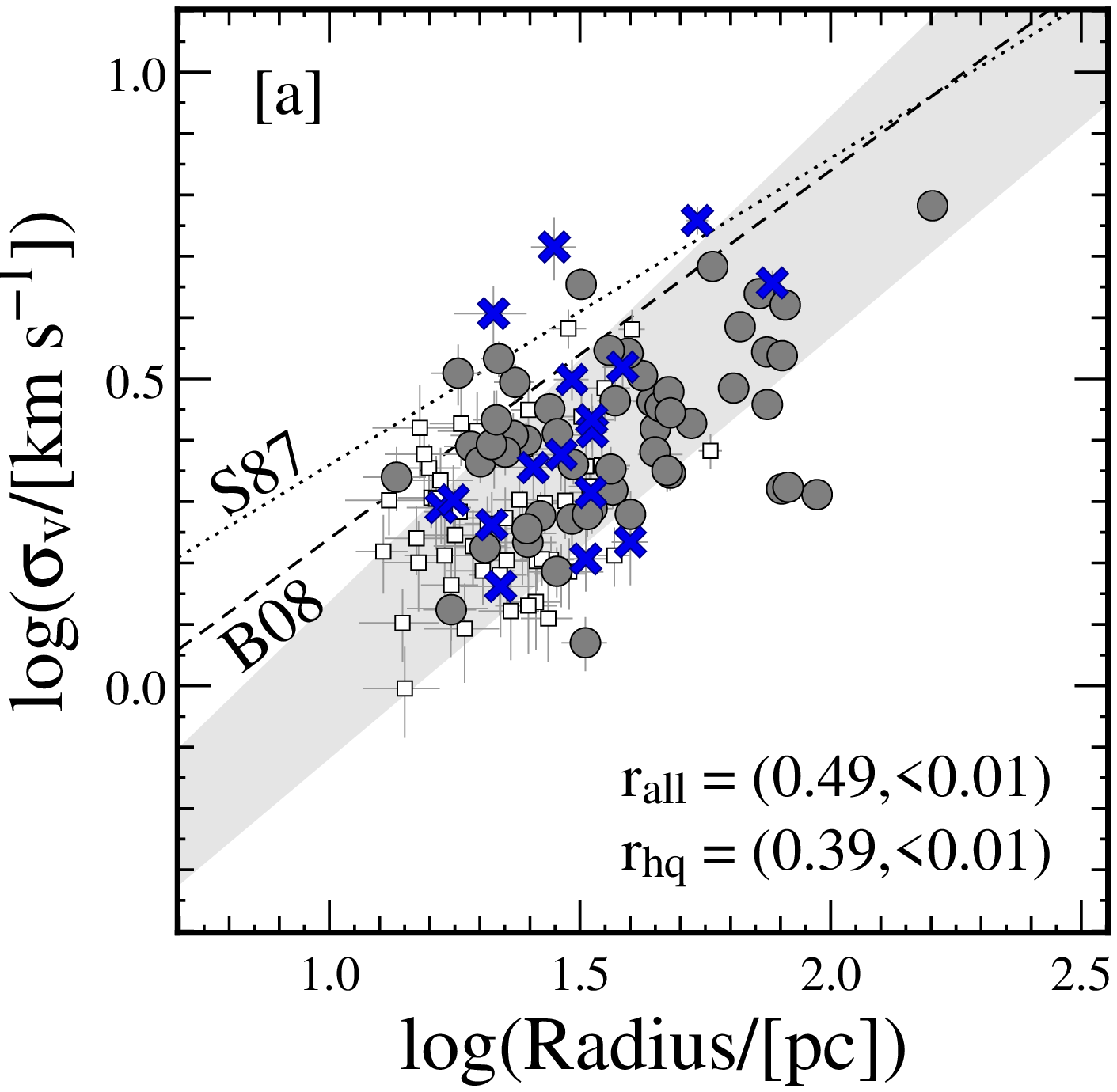}
\hspace{0.5cm}
\includegraphics[width=70mm]{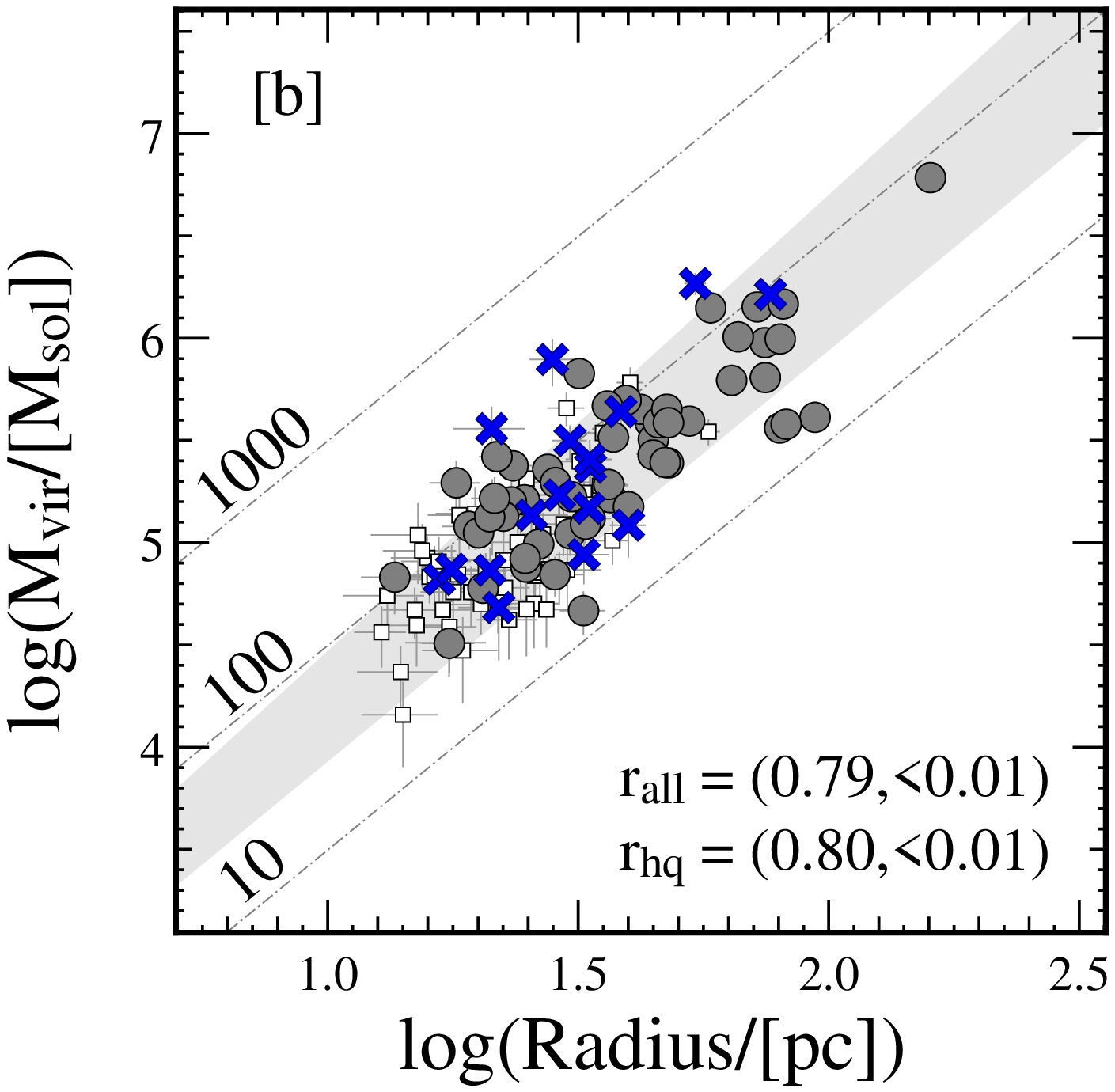}
\includegraphics[width=70mm]{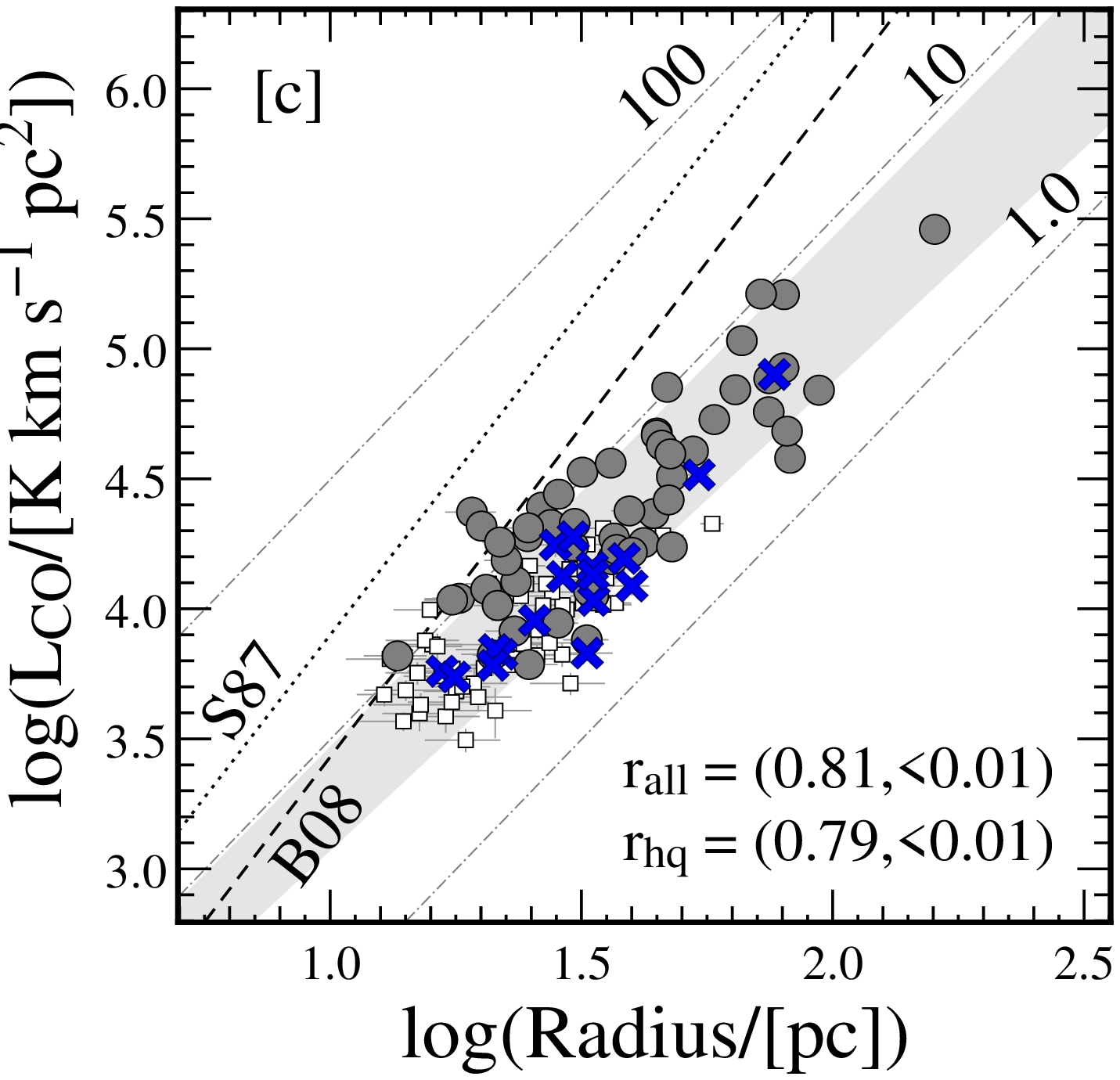}
\hspace{0.5cm}
\includegraphics[width=70mm]{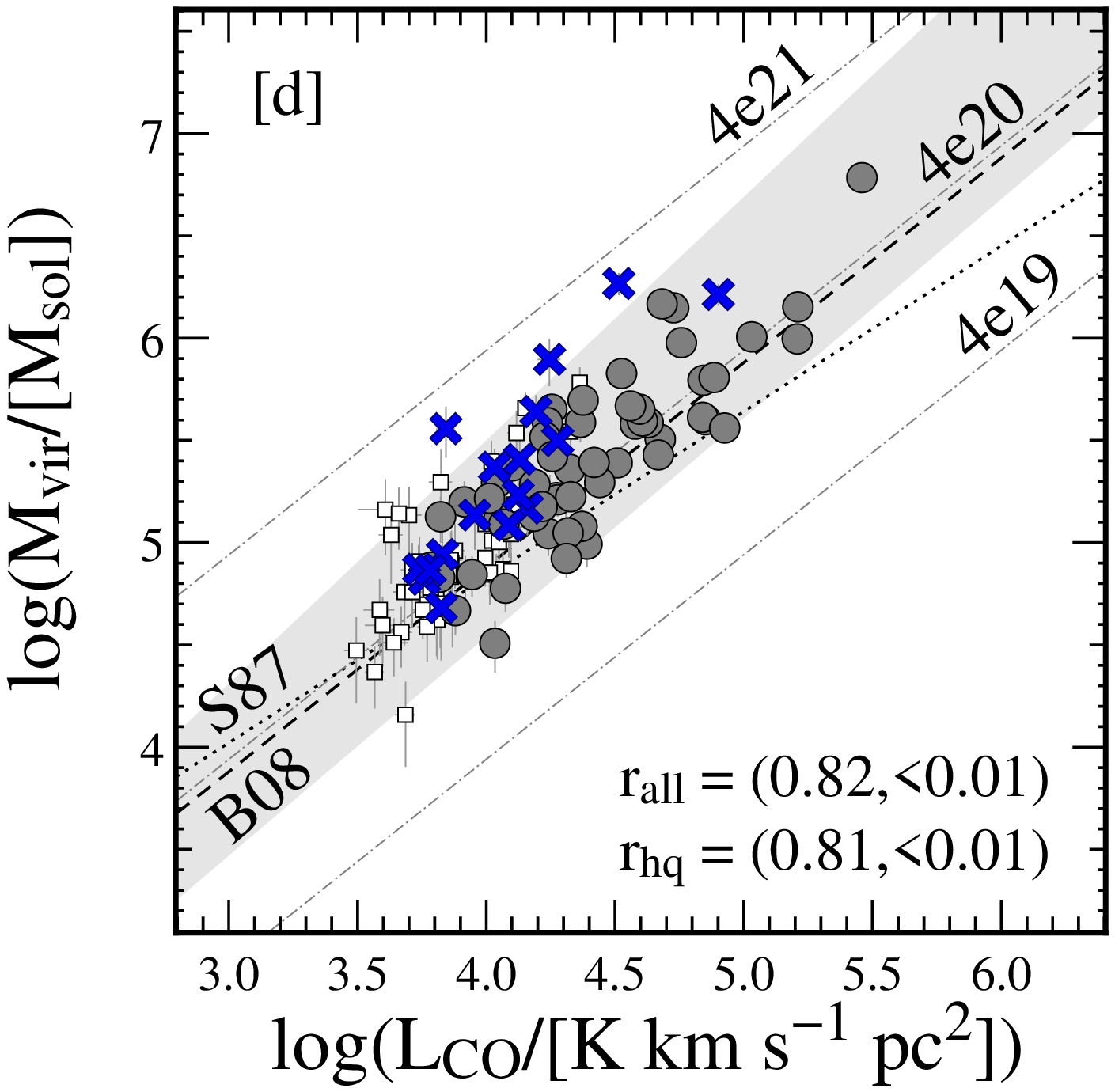}

\caption{Plots of (a) radius versus velocity dispersion; (b) radius
  versus virial mass; (c) radius versus CO luminosity; and (d) CO
  luminosity versus virial mass for the GMCs identified in the MAGMA
  LMC survey. In each panel, the light grey shaded area represents our
  BCES bisector fit to the 125 GMCs in the MAGMA cloud list and the
  1-$\sigma$ uncertainty in this fit (see text). The black dotted line
  in each panel shows the standard relation for the S87 inner Milky
  Way data, while the black dashed line represents the relation
  determined for extragalactic GMCs by B08. The dot-dashed grey lines
  represent constant values of CO surface brightness ($I_{\rm CO} =
  1$, 10, 100~\kkms, panel [b]); mass surface density ($\Sigma_{\rm
    H_{2}} = 10$, 100, 1000~\mpcsq, panel [c]); and
  CO-to-\hh\ conversion factor ($\xco = 0.4, 4.0, 40 \times
  10^{20}~\xcou$, panel [d]). GMCs belonging to the complete MAGMA LMC
  cloud list are represented by small open squares, and GMCs in the
  high quality subsample are indicated using filled grey circles. The
  blue cross symbols represent GMCs without signs of active star
  formation. For comparison with the correlations presented in
  Fig.~\ref{fig:isrfcmp} to~\ref{fig:phcmp}, Spearman's rank
  correlation coefficent and corresponding $p$-value for all clouds
  and the high quality subsample are indicated at the bottom right of
  each panel.}
\label{fig:Rdv}
\end{center}
\end{figure*}

\noindent Fig.~\ref{fig:Rdv}[a] presents a plot of velocity dispersion
versus radius for GMCs in the MAGMA cloud list. The LMC clouds are
clearly offset towards lower velocity dispersions compared to the
$R-\sigma_{\rm v}$ relations derived by S87 and B08; some of the
larger LMC clouds fall under the $R-\sigma_{\rm v}$ relation
determined by B08 ($\sigma_{\rm v} \approx 0.44^{+0.18}_{-0.13}
R^{0.6\pm0.1}$) by a factor of $\sim3$ in velocity dispersion. The
discrepancy between the LMC clouds and the $R-\sigma_{\rm v}$ relation
derived by B08 should not be the result of cloud decomposition
techniques, as we have identified cloud and parameterized GMC
properties using the same algorithms as these authors. In the
turbulent paradigm, the offset towards lower velocity dispersion at a
given radius would suggest that the turbulent bulk motions within LMC
molecular clouds are more quiescent than in the B08 GMCs. If, on the
other hand, GMCs in the LMC manage to achieve rough dynamic
equilibrium, then their relatively narrow linewidths would imply that
they have lower mass surface densities than the B08 clouds.  \\

\noindent To fit the $R-\sigma_{\rm v}$ relationship for the MAGMA
data, we used the BCES bisector linear regression method presented by
\citet[][]{akritasbershady96}, which is designed to take measurement
errors in both the dependent and independent variable, and the
intrinsic scatter of a dataset into account.\footnote{ 'BCES' stands
  for bivariate, correlated errors and intrinsic scatter. Software
  that implements this method is available from: {\tt
    http://www.astro.wisc.edu/$\sim$mab/archive/stats/stats.html}. For
  our analysis, we assume that measurement errors are uncorrelated.}
For our analysis of all the Larson relations, we use the bisector
method because our goal is to estimate the intrinsic relation between
the cloud properties \citep[e.g.][]{babufeigelson96}. The best-fitting
relation for all 125 clouds in the MAGMA cloud list, illustrated with
grey shading in Fig.~\ref{fig:Rdv}[a], is:
\begin{equation}
\log \sigma_{\rm v} = (-0.73 \pm 0.08) + (0.74 \pm 0.05) \log R \textrm{  (BCES bisector)}.
\end{equation}
\noindent Here, and for all other relations presented in this paper,
the errors in the regression coefficients are derived using
bootstrapping techniques; for relations determined using a BCES
estimator, they are consistent with the standard deviation of the
regression coefficients derived according to eqn.~30 in
\citet{akritasbershady96} unless otherwise noted. Although the errors
in best-fitting relation that we derive using are small, we caution
that the form of the relation depends on the linear regression method
\citep[for a discussion of this issue in other astronomical contexts,
  see e.g.][]{tremaineetal02,kelly07,blancetal09}. For comparison with
other work, the best-fitting relations determined using the BCES
ordinary least squares (henceforth OLS(Y$|$X)) and BCES orthogonal
methods are:
\begin{equation}
\log \sigma_{\rm v} = (-0.24 \pm 0.09) + (0.40 \pm 0.06) \log R \textrm{  (BCES OLS(Y|X))}
\end{equation}
\begin{equation}
\log \sigma_{\rm v} = (-0.46 \pm 0.12) + (0.56 \pm 0.08) \log R \textrm{  (BCES orthogonal)}.
\end{equation}
\noindent The best-fitting relation derived using the FITEXY estimator
\citep{pressetal92} is:
\begin{equation}
\log \sigma_{\rm v} = (-0.37 \pm 0.17) + (0.50 \pm 0.11) \log R \textrm{  (FITEXY)}.
\end{equation}

\noindent The intrinsic scatter of the MAGMA GMCs around the
best-fitting $R-\sigma_{\rm v}$ relation is greater than the
measurement errors in $R$ and $\sigma_{\rm v}$ across the observed
range of cloud radii. The $C_{0}$ values of individual MAGMA GMCs vary
between 0.21 and 0.98~\kms~pc$^{-0.5}$. Although LMC clouds may be
said to follow the same $R-\sigma_{\rm v}$ relation as other
extragalactic GMCs in the sense that the slope and amplitude of the
derived best-fitting relations are similar, their molecular mass
surface density is not strictly constant, but instead varies between
15 and 320~\mpcsq\ (see Fig.~\ref{fig:Rdv}[b]). An outstanding
question, which we investigate further in Section~\ref{sect:tracers},
is whether the variation in $\Sigma_{\rm H_{2}}$ is stochastic or
whether it is related to changes in the environment of the GMCs.

\subsection{The size-luminosity relation}
\label{sect:RL}

\noindent Fig.~\ref{fig:Rdv}[c] presents a plot of CO luminosity
versus radius for GMCs in the MAGMA cloud list. Although
the $L_{\rm CO}-R$ relation determined for extragalactic clouds by B08
($L_{\rm CO} \approx 7.8^{+6.9}_{-3.7}R^{2.54\pm0.20}$) overlaps with
the smaller MAGMA clouds, the slope of the $L_{\rm CO}-R$ relation in
the LMC is clearly shallower, such that the CO emission in large LMC
molecular clouds is up to an order of magnitude fainter than for GMCs
in nearby galaxies. A BCES bisector fit yields:
\begin{equation}
\log L_{\rm CO} = (1.39 \pm 0.12) + (1.88 \pm 0.08) \log R,
\end{equation}
for the $L_{\rm CO}-R$ relation. The relations determined using the
BCES OLS(Y$|$X), BCES orthogonal and FITEXY methods are also shallower
than the B08 fit, with slopes of 1.67$\pm$0.09, 2.02$\pm$0.10 and
2.17$\pm$0.29 respectively. \\

\noindent The median CO surface brightness of GMCs in the MAGMA cloud
list is 4.8~\kkms, and the median absolute deviation of the GMCs
around this value is $\sim30$ per cent. This corresponds to an average
mass surface density of only $\langle \Sigma_{H_{2}} \rangle \sim
20$~\mpcsq\ for the MAGMA clouds if we adopt the Galactic \xco\ value
\citep[$2 \times 10^{20}$~\xcou, e.g.][]{strongmattox96} and $\langle
\Sigma_{H_{2}} \rangle \sim 50$~\mpcsq\ if we use the median
\xco\ value of the MAGMA cloud list ($\xco = 4.7 \times
10^{20}$~\xcou, see Table~\ref{tbl:ks1}). Individual GMCs have $I_{\rm
  CO}$ values between 1.8 and 20.4~\kkms. Some of the variation in
$I_{\rm CO}$ appears to be related to the location of the GMCs: for
clouds that are coincident with the stellar bar (defined as regions
where $\Sigma_{*} > 100$~\mpcsq, see Fig.~\ref{fig:map}), the median
CO surface brightness is 9.1~K~\kms. We examine the relationship
between the CO emission in LMC molecular clouds and the stellar mass
surface density more closely in Section~\ref{sect:cmpstellar}.

\subsection{The calibration between virial mass and CO luminosity}
\label{sect:LMvir}

\noindent Fig.~\ref{fig:Rdv}[d] presents a plot of the virial mass
estimate versus the CO luminosity for the MAGMA clouds. The LMC clouds
appear to be in reasonable agreement with the slope of the relation
determined for the B08 extragalactic GMC data ($M_{\rm vir} \approx
7.6^{+3.2}_{-2.6} L_{\rm CO}^{1.00\pm0.04}$), although offset slightly
to higher $M_{\rm vir}$ values. A BCES bisector fit to the $M_{\rm
  vir}-L_{\rm CO}$ relation for the complete MAGMA LMC cloud list yields:
\begin{equation}
\log M_{\rm vir} = (0.50 \pm 0.25) + (1.13\pm 0.06) \log L_{\rm CO}.
\end{equation}
\noindent The slopes of the relations determined using the BCES
OLS(Y$|$X), BCES orthogonal and FITEXY methods are 0.99$\pm$0.06,
1.16$\pm$0.07 and 0.97$\pm$0.17 respectively. Even for the steepest
BCES fit, the systematic variation in \xco\ with mass is small,
corresponding to only a factor of two increase in \xco\ for GMC masses
between $3 \times 10^{4}$ and $3 \times 10^{6}$~\msol. The rms scatter
of the MAGMA GMCs around the best-fitting relation corresponds to a
similar variation in \xco.\\

\noindent The median value of \xco\ for the MAGMA LMC clouds is $4.7
\times 10^{20}$~\xcou. This is in excellent agreement with the values
derived by B08 and \citet{blitzetal07} for their extragalactic GMCs,
and the LMC value obtained by the SEST Large Programme
\citep{israeletal03}, but lower than the value derived from the NANTEN
LMC survey \citep[$7 \times 10^{20}$~\xcou,][]{fukuietal08}. While the
{\textsc CPROPS} cloud decomposition algorithm aims to minimize
instrumental effects, part of this discrepancy may be due to the
difference in angular resolution between the two surveys. GMCs are
constituted by dense ($n \approx 10^{3}$~\ccc) CO-bright peaks
embedded in more diffuse gas with lower CO brightness; as noted by
several previous authors \citep[e.g.][]{pinedaetal09,bolattoetal03},
observations with coarser resolution trace larger physical structures
and hence derive larger values for \xco. In principle, the discrepancy
could also arise from MAGMA's observational strategy, which excluded
clouds in the NANTEN catalogue with low CO surface brightness (recall
that $\xco \equiv \Sigma_{\rm H_{2}}/2.2 I_{\rm CO}$). In practice,
however, we do not expect that our target selection has a significant
impact on the average MAGMA \xco\ value, since there are only four
NANTEN clouds with $L_{\rm CO} >7000$~\kkms~pc$^{2}$ and peak $I_{\rm
  CO} < 1~\kkms$. 

\section{GMC properties and interstellar conditions}
\label{sect:tracers}

\noindent In this section, we explore the variation of the physical
properties of the MAGMA LMC clouds in response to local interstellar
conditions. We measure the local \hi\ column density, stellar mass
surface density, interstellar radiation field and external pressure
for each GMC using the maps described in Section~\ref{sect:ancillary},
taking the mean value of all independent pixels with integrated CO
emission greater than 1~\kkms. We measure the strength of
correlations between the GMC and interstellar properties using the
Spearman rank correlation coefficient, $r$, a non-parametric rank
statistic that measures the strength of monotone association between
two variables. The statistical significance of $r$ is assessed by
calculating the corresponding $p$-value, which is the two-sided
significance level of $r$'s deviation from zero. We consider
$p$-values less than 0.01 to provide statistically significant
evidence against the null hypothesis (i.e. that there is no underlying
correlation between the variables). We regard $|r|$ values greater
than 0.6 as strong correlations (or anti-correlations if $r < 0)$,
$|r|$ values between 0.4 and 0.6 as moderate correlations, and $|r|$
values between 0.2 and 0.4 as weak correlations.  We regard
correlations with $|r|$ values less than 0.2 to be very weak and
therefore unlikely to have practical significance, even if their
$p$-value is small.\\

\noindent Each correlation test was repeated for the complete MAGMA
LMC cloud list and for the high quality subsample. Since the Spearman
rank correlation test does not account for measurement uncertainties,
we performed 500 trials of each correlation test in which we offset
each cloud property measurement by a fraction of its uncertainty. As
for the KS test in Section~\ref{sect:younggmcs}, the value of a cloud
property measurement in each trial was displaced by $k\Delta x$, where
$\Delta x$ is the absolute uncertainty in the cloud property
measurement and $k$ is a uniformly distributed random number between
-1 and 1. We consider our correlation result to be robust and
significant if i) the correlation coefficients obtained in the trials
are narrowly distributed around the original (i.e. unperturbed) value
of $r$, ii) the corresponding $p$-values are narrowly distributed
around zero, i.e. $\langle p \rangle \leq 0.01$ and $\sigma_{p} \leq
0.01$, and iii) $|r| \geq 0.2$ for both the complete MAGMA cloud list
and the high quality subsample. The results of all the correlation
tests are presented in Table~\ref{tbl:spearman}. As a reference for
the values obtained in these comparisons, we conducted equivalent
correlation tests for the Larson scaling relations shown in
Fig.~\ref{fig:Rdv}, and present the results in
Table~\ref{tbl:spear_larson}. We note that the size-linewidth relation
is only a weak to moderate correlation: for all 125 GMCs in the MAGMA
cloud list $\langle r \rangle = 0.48$, while for the high quality GMCs
$\langle r \rangle = 0.39$. The size-luminosity and mass-luminosity
relations are more strongly correlated ($\langle r \rangle \approx
0.8$); this is expected since  $L_{\rm CO}$ and $M_{\rm vir}$ are both
dependent on $R$ and $\sigma_{\rm v}$.\\

\begin{table*}
\begin{minipage}{170mm}
\caption{Results of the correlation tests between the properties of
  the MAGMA GMCs and properties of the interstellar environment. The
  properties used in the comparison are listed in columns 1 and 2.
  The results of the correlation tests for the complete MAGMA cloud
  list are shown in columns 3 to 5; the results for the 57 high quality
  GMCs are shown in columns 6 to 8. The results list 
  $\langle r \rangle$, $\langle p \rangle$ and $\sigma_{p}$,
  where $\langle r \rangle$ is the median Spearman correlation coefficient 
  obtained in the error trials (see text), $\langle p \rangle$ is the median 
  of the corresponding $p$-values, and $\sigma_{p}$ is the standard deviation
 of the $p$-values.}
\label{tbl:spearman}
\begin{tabular}{@{}llcccccc}
\hline
                &               & \multicolumn{3}{|c|}{All GMCs}                        & \multicolumn{3}{|c|}{HQ GMCs} \\
Environment  	& GMC 		&  $\langle r \rangle$ & $\langle P \rangle$ & $\sigma(P)$ & $\langle r \rangle$ & $\langle P \rangle$ & $\sigma(P)$  \\
\hline
$G_{0}$	 & $R$		        &    -0.16  &  0.06  &  0.03    	&   -0.27  &  0.03  &  0.01  \\
	& $\sigma_{\rm v}$	&    -0.03  &  0.71  &  0.16 	&   0.13  &  0.39  &  0.11 	\\
	& $\langle T_{\rm pk} \rangle$ &   0.18  &  0.04  &  0.03 	&   0.26  &  0.05  &    0.03   \\
	& $I_{\rm CO}$		        &    0.17  &  0.06  &  0.03 	&    0.38  &  $<0.01$  &  $<0.01$ \\
	& $I_{\rm CO}$, SF only		&    0.13  &  0.17  &  0.08 	&    0.38  &  $<0.01$  &  $<0.01$ \\
	& $\Sigma_{\rm H_{2}}$	        &     0.06  &  0.52  &  0.20 	&   0.27  &  0.04  &  0.03  \\
	& $\Sigma_{\rm H_{2}}$, SF only	&     0.17  &  0.08  &  0.05 	&  0.37  &  $<0.01$  &  $<0.01$  \\
	& \xco\		                &     -0.06  &  0.51  &  0.21 	&    -0.03  &  0.83  &  0.14 \\
\hline
$\Sigma_{*}$	 & $R$		&     -0.13  &  0.13  &  0.05 	&   -0.10  &  0.36  &  0.08  \\
	& $\sigma_{\rm v}$	&     0.00  &  0.87  &  0.10 	&   0.09  &  0.54  &  0.13   \\
	& $\langle T_{\rm pk} \rangle$  &   0.34  &  $<0.01$  &  $<0.01$   &  0.55  &  $<0.01$  &  $<0.01$  \\
	& $I_{\rm CO}$		&     0.34  &  $<0.01$  &  $<0.01$   &  0.40  &  $<0.01$  &  $<0.01$  \\
	 & $\Sigma_{\rm H_{2}}$	&    0.06  &  0.44  &  0.19 	&   0.10  &  0.41  &  0.14  \\
	& \xco	                &    -0.19  &  0.04  &  0.03 	&    -0.22  &  0.08  &  0.04  \\
\hline
$N \hi $	 & $R$		&    0.12  &  0.19  &  0.05      &   0.28  &  0.04  &  0.01  \\
	& $\sigma_{\rm v}$	&   0.30  &  $<0.01$  &  $<0.01$     &  0.43  &  $<0.01$  &  $<0.01$   \\
	& $\langle T_{\rm pk} \rangle$  &  -0.07  &  0.36  &  0.15 	&  -0.15  &  0.24  &  0.10  \\
	& $I_{\rm CO}$		&   -0.05  &  0.58  &  0.14 	&  -0.03  &  0.90  &  0.08  \\
	& $\Sigma_{\rm H_{2}}$	&   0.25  &  $<0.01$  &  $<0.01$    &  0.26  &  0.06  &  0.03  \\
	& \xco		 &   0.28  &  $<0.01$  &  $<0.01$  &   0.29  &  0.03  &  0.02  \\
	& \xco, SF only	  &  0.32  &  $<0.01$  &  $<0.01$  &   0.28  &  0.04  &  0.02  \\
\hline
$P_{h}$	 & $R$		 &   0.03  &  0.76  &  0.15 	&    0.19  &  0.15  &  0.05  \\
	& $\sigma_{\rm v}$	&   0.32  &  $<0.01$  &  $<0.01$  &   0.49  &  $<0.01$  &  $<0.01$ \\
	& $\langle T_{\rm pk} \rangle$  &  0.13  &  0.16  &  0.09 	&   0.25  &  0.07  &  0.04  \\
	& $I_{\rm CO}$		&    0.16  &  0.07  &  0.03 	&   0.25  &  0.06  &  0.03  \\
	& $\Sigma_{\rm H_{2}}$	&    0.27  &  $<0.01$  &  $<0.01$  &   0.34  &  0.01  &  0.01  \\
	& \xco		 &   0.14  &  0.13  &  0.08 	&  0.14  &  0.32  &  0.12  \\

\hline
\end{tabular}
\end{minipage}
\end{table*}

\begin{table*}
\begin{minipage}{170mm}
\caption{Results of the correlation tests for Larson's scaling
  relations in the LMC. The relation is listed in column 1. The results 
  of the correlation tests for the complete MAGMA cloud
  list are shown in column 2; the results for the 57 high quality
  GMCs are shown in column 3. The results are expressed in the form 
  ($\langle r \rangle$, $\langle p \rangle$, $\sigma_{p}$),
  where $\langle r \rangle$ is the median Spearman correlation coefficient 
  obtained in the error trials (see text), $\langle p \rangle$ is the median 
  of the corresponding $p$-values, and $\sigma_{p}$ is the standard deviation
 of the $p$-values.}
\label{tbl:spear_larson}
\begin{tabular}{@{}lcccccc}
\hline
                        & \multicolumn{3}{|c|}{All GMCs}                        & \multicolumn{3}{|c|}{HQ GMCs} \\
Relation            	&  $\langle r \rangle$ & $\langle P \rangle$ & $\sigma(P)$ & $\langle r \rangle$ & $\langle P \rangle$ & $\sigma(P)$  \\ 
\hline
$R$ - $\sigma_{\rm v}$		 & 0.48  &   $<0.01$  &   $<0.01$  	& 0.39  &  $<0.01$  &  $<0.01$  \\			
$R$ - $L_{\rm CO}$		 & 0.80  &  $<0.01$  &  $<0.01$  		                & 0.80  &  $<0.01$  &  $<0.01$   \\             
$L_{\rm CO}$ - $M_{\rm vir}$	  & 0.81  &  $<0.01$  &  $<0.01$  		        & 0.80  &  $<0.01$  &  $<0.01$	\\                    
\hline
\end{tabular}
\end{minipage}
\end{table*}

\subsection{Comparison with $G_{0}$}
\label{sect:cmpisrf}

\begin{figure*}
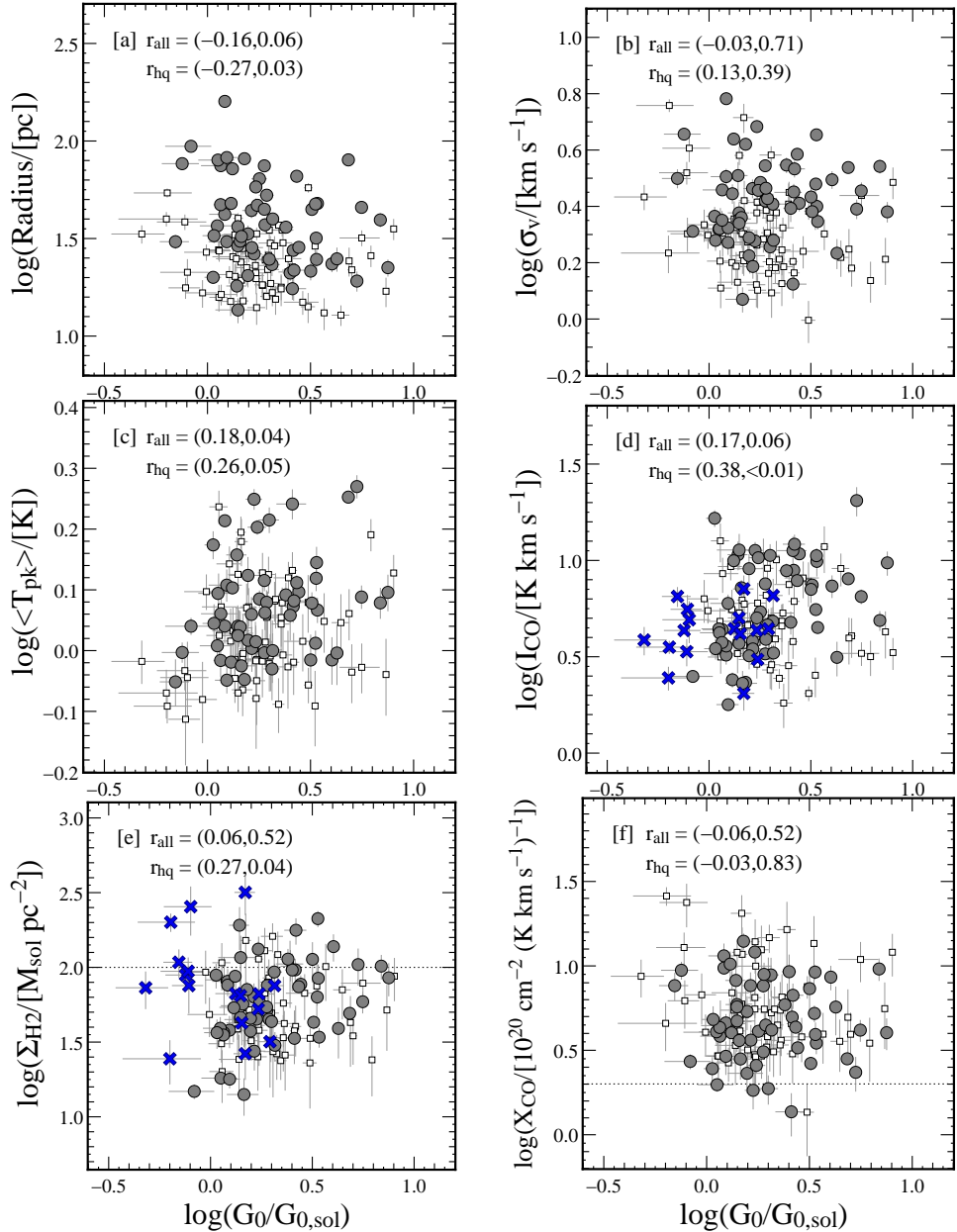

\begin{center}
\includegraphics[width=60mm,angle=0]{fig5a.epsi}
\hspace{0.5cm}
\includegraphics[width=60mm,angle=0]{fig5b.epsi}
\includegraphics[width=60mm,angle=0]{fig5c.epsi}
\hspace{0.5cm}
\includegraphics[width=60mm,angle=0]{fig5d.epsi}
\includegraphics[width=60mm,angle=0]{fig5e.epsi}
\hspace{0.5cm}
\includegraphics[width=60mm,angle=0]{fig5f.epsi}
\caption{Properties of the MAGMA GMCs, plotted as a function of the
  local interstellar radiation field, $G_{0}/G_{0,\odot}$: (a) radius;
  (b) velocity dispersion; (c) average peak CO temperature; (d) CO
  surface brightness, $I_{\rm CO}$; (e) molecular mass surface
  density, $\Sigma_{\rm H_{2}}$; and (f) CO-to-\hh\ conversion factor,
  \xco. The plot symbols are the same as in Fig.~\ref{fig:Rdv}. The
  horizontal dotted lines in panels [e] and [f] indicate $\Sigma_{\rm
    H_{2}} = 100$~\mpcsq\ and $\xco \sim 2 \times 10^{20}$~\xcou,
  i.e. values that apply to GMCs in the inner Milky Way. The Spearman
  rank correlation coeffient and corresponding $p$-value for the
  complete MAGMA cloud list and high quality subsample are indicated
  at the top left of each panel.}
\label{fig:isrfcmp}
\end{center}
\end{figure*}

\noindent In Fig.~\ref{fig:isrfcmp}, we plot the radius, velocity
dispersion, average peak CO brightness, CO surface brightness, mass
surface density and CO-to-\hh\ conversion factor of the MAGMA clouds
as a function of the local interstellar radiation field, $G_{0}$. All
the plots show considerable scatter and none of the correlations
satisfy our criteria for significance. Contrary to what might be
expected from classic photodissociation models
\citep[e.g.][]{vandishoeckblack88}, there is no general trend between
the $G_{0}$ and \xco, even for clouds without signs of active star
formation (i.e. where $G_{0}$ is dominated by the external field). An
earlier analysis of the MAGMA data for the molecular ridge region also
found that \xco\ was insensitive to variations in the radiation field
strength \citep{pinedaetal09}. \\

\subsection{Comparison with the interstellar pressure}
\label{sect:cmpph}

\noindent In this subsection, we investigate whether the physical
properties of the MAGMA clouds vary with the interstellar pressure. We
estimate the total pressure at the molecular cloud boundary using the
expression given by E89 for a two-component disc of gas and stars in
hydrostatic equilibrium:
\begin{equation}
P_{h} = \frac{\pi G}{2}\Sigma_{g}\left(\Sigma_{g} + \frac{\sigma_{g}}{\sigma_{*}}\Sigma_{*}\right).
\label{eqn:pressure}
\end{equation}
\noindent In this expression, $\sigma_{g}$ and $\sigma_{*}$ are the
velocity dispersions of the gas and stars respectively, $\Sigma_{g}$
is the mass surface density of the gas and $\Sigma_{*}$ is the stellar
mass surface density. The term in brackets on the right hand side of
Equation~\ref{eqn:pressure} is an estimate for the total dynamical
mass within the disc gas layer. We assume a constant velocity
dispersion of $\sigma_{g} = 9$~\kms\ for the gas, based on the average
dispersion of the \hi\ line profiles across the LMC \citep[see
  also][]{wongetal09}, and $\sigma_{*} = 20$~\kms\ for the stars
\citep{vandermareletal02}. For $\Sigma_{*}$, we use the stellar mass
surface density map of \citet{yangetal07} described above, and we
estimate $\Sigma_{g} \equiv \Sigma_{\hi} [\mpcsq] = 1.089 \times
10^{-20}\nh [\cc]$ directly from the \hi\ column density map (the
conversion between \nh\ and $\Sigma_{g}$ includes a factor of 1.36 by
mass for the presence of helium). We note that the \hi\ and stars make
similar contributions to the total mass surface density at the
locations of GMCs in the LMC: $\langle \Sigma_{*} \rangle =
48$~\mpcsq, and $\langle \Sigma_{g} \rangle = 30$~\mpcsq; for
comparison, E89 adopted $\langle \Sigma_{*} \rangle = 55$~\mpcsq\ and
$\langle \Sigma_{g} \rangle = 12$~\mpcsq\ as characteristic values in
the Milky Way disc.\\

\noindent Along sightlines to GMCs, our definition of $P_{h}$ provides
an estimate of the pressure at the surface of the molecular cloud
rather than at the disc midplane, since the weight of a GMC is likely
to make a significant contribution to the total midplane
pressure. Further caveats are (i) the assumption that the disc is
supported against gravity solely by the observed velocity dispersions,
and (ii) that the interaction between the LMC and the Galactic halo
may make a significant contribution to the pressure that is not
accounted for by our simple estimate in
Equation~\ref{eqn:pressure}. In the following, we first compare the
cloud properties to two independent components of $P_{h}$ for which we
have an empirical tracer: i) the stellar mass surface density,
$\Sigma_{*}$ and ii) the atomic gas surface density, $\Sigma_{g}$.

\subsubsection{Comparison with $\Sigma_{*}$}
\label{sect:cmpstellar}

\begin{figure*}
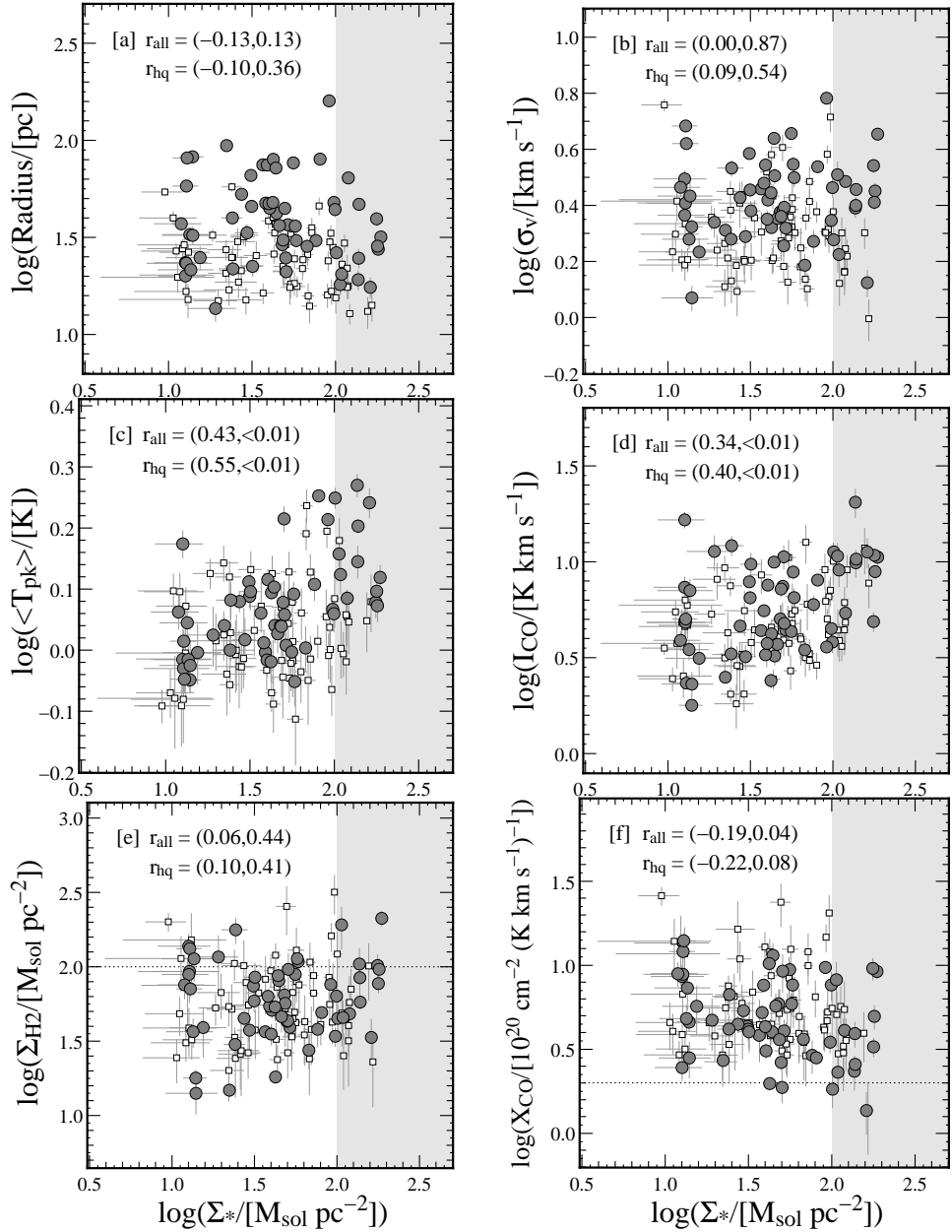

\begin{center}
\includegraphics[width=60mm,angle=0]{fig6a.epsi}
\hspace{0.5cm}
\includegraphics[width=60mm,angle=0]{fig6b.epsi}
\includegraphics[width=60mm,angle=0]{fig6c.epsi}
\hspace{0.5cm}
\includegraphics[width=60mm,angle=0]{fig6d.epsi}
\includegraphics[width=60mm,angle=0]{fig6e.epsi}
\hspace{0.5cm}
\includegraphics[width=60mm,angle=0]{fig6f.epsi}
\caption{Properties of the MAGMA clouds, plotted as a function of the
  stellar mass surface density, $\Sigma_{*}$. The panels, plot symbols
  and annotations are the same as in Fig.~\ref{fig:isrfcmp}. The grey
  shading indicates regions where $\Sigma_{*} \geq 100$~\mpcsq,
  corresponding to the stellar bar region denoted by the ellipse in
  Fig.~\ref{fig:map}.}
\label{fig:stellarcmp}
\end{center}
\end{figure*}

\noindent In Fig.~\ref{fig:stellarcmp}, we plot the properties of the
MAGMA clouds versus $\Sigma_{*}$. The correlation tests indicate weak
but significant correlations between $\Sigma_{*}$ and $\langle T_{\rm
  pk} \rangle$ (panel [c]) and $I_{\rm CO}$ (panel [d]); these
correlations are strengthened if we restrict our analysis to high
quality GMCs. The existence of these trends suggests that the GMCs in
the LMC do respond to the presence of the galaxy's stellar population,
despite evidence that the stellar bar may be physically offset from
the gas disc \citep[e.g.][]{zhaoevans00,nikolaevetal04}.

\subsubsection{Comparison with $\Sigma_{g}$}
\label{sect:cmphi}

\begin{figure*}
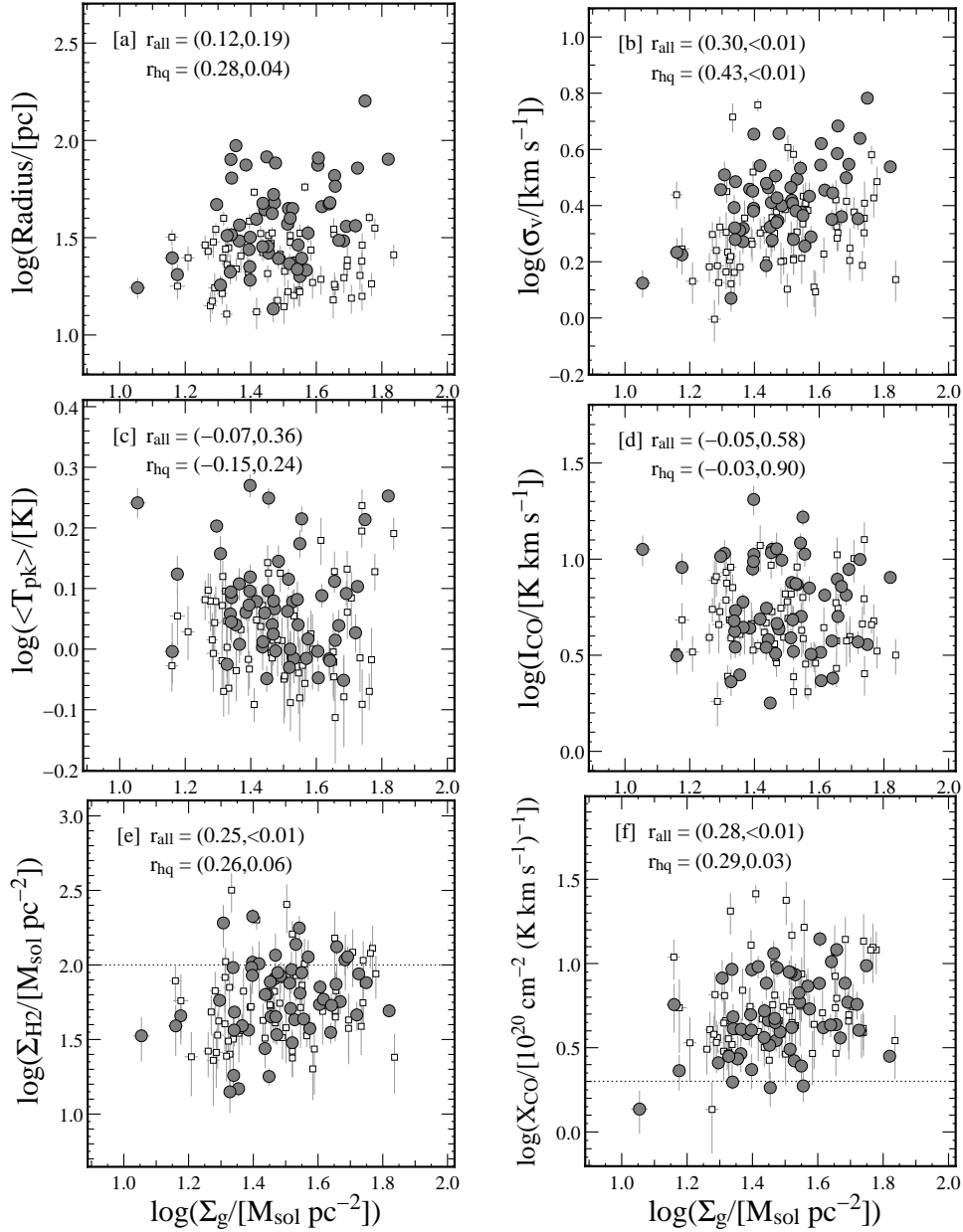

\begin{center}
\includegraphics[width=60mm,angle=0]{fig7a.epsi}
\hspace{0.5cm}
\includegraphics[width=60mm,angle=0]{fig7b.epsi}
\includegraphics[width=60mm,angle=0]{fig7c.epsi}
\hspace{0.5cm}
\includegraphics[width=60mm,angle=0]{fig7d.epsi}
\includegraphics[width=60mm,angle=0]{fig7e.epsi}
\hspace{0.5cm}
\includegraphics[width=60mm,angle=0]{fig7f.epsi}
\caption{Properties of the MAGMA clouds, plotted as a function of the
  atomic gas surface density, $\Sigma_{g}$. The panels, plot symbols
  and annotations are the same as in Fig.~\ref{fig:isrfcmp}.}
\label{fig:nhcmp}
\end{center}
\end{figure*}

\noindent In Fig.~\ref{fig:nhcmp}, we plot the cloud properties as a
function of the atomic gas surface density, $\Sigma_{g}$, which is
estimated from the total \hi\ column density along the line-of-sight.
We find a weak but robust correlation between $\Sigma_{g}$ and
$\sigma_{\rm v}$ (panel [b]); again, the correlation is stronger if
only high quality GMCs are considered. There is some indication that
$\Sigma_{\rm H_{2}}$ (panel [e]) and \xco\ (panel [f]) also increase
with the \hi\ column density -- as would be expected if $\sigma_{\rm
  v}$ increases without a corresponding increase in cloud size or CO
luminosity -- but the correlations are not significant if only high
quality GMCs are considered. An important caveat for interpreting
Fig.~\ref{fig:nhcmp} is that \hi\ line profiles in the LMC are
notoriously complex, exhibiting two well-defined velocity components
in some parts of the LMC's disc \citep[especially in the south-east,
  e.g.][]{luksrohlfs92}. The CO emission is almost invariably
associated with only one \hi\ velocity component \citep{wongetal09},
so the \hi\ column density that is physically associated with a GMC is
almost certainly overestimated by the total \nh\ along the
line-of-sight in these regions. Excluding GMCs located in the
south-east of the LMC (i.e. clouds with right ascension above 05h38m
and declination below -68d30m (J2000)) from our comparisons between
the cloud properties and $\Sigma_{g}$ suggests that the correlations
between $\Sigma_{g}$ and $\sigma_{v}$ is not severely contaminated by
the contribution of the secondary \hi\ velocity component to \nh.

\subsubsection{Comparison with $P_{h}$}
\label{sect:cmpph2}

\begin{figure*}
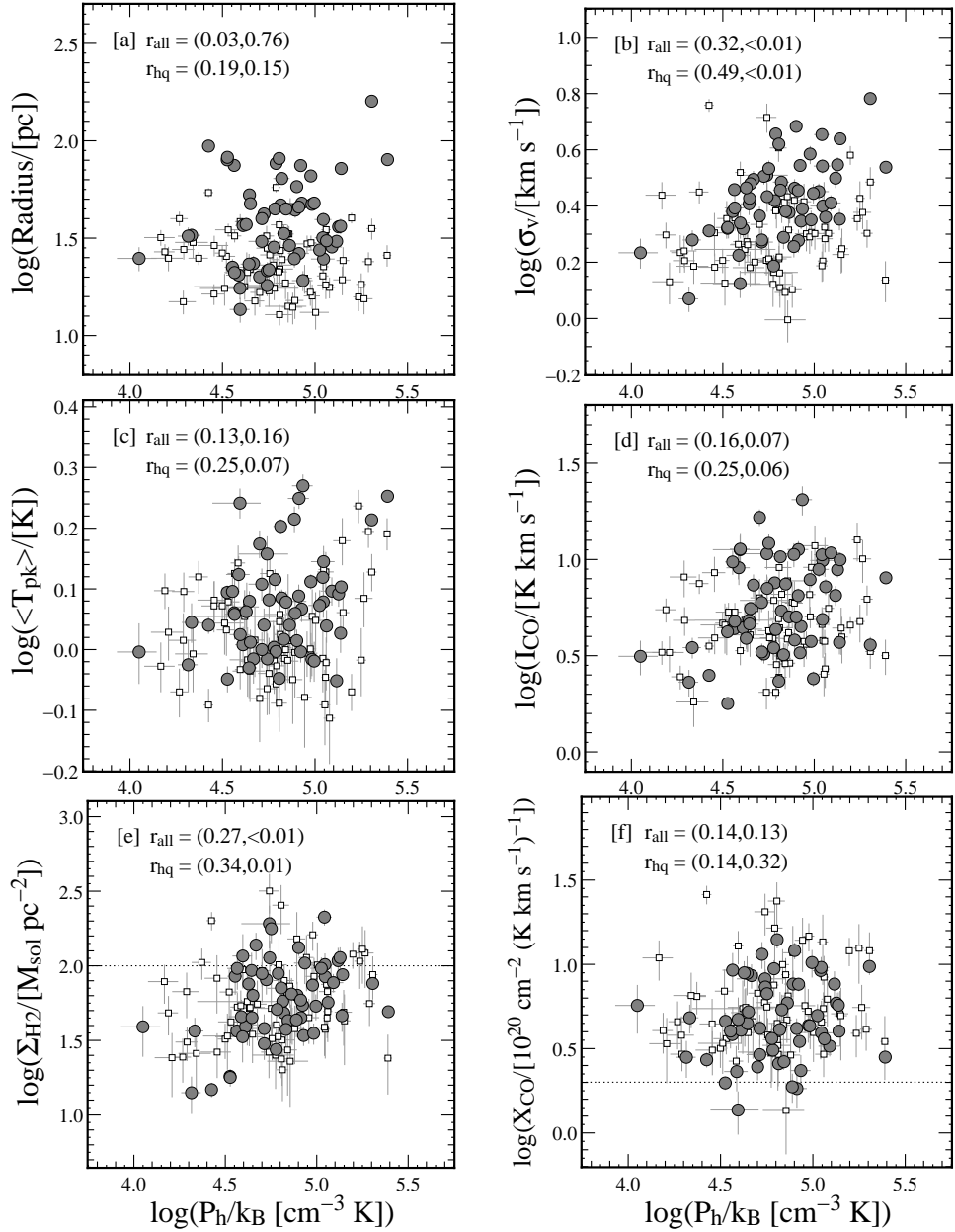

\begin{center}
\includegraphics[width=60mm,angle=0]{fig8a.epsi}
\hspace{0.5cm}
\includegraphics[width=60mm,angle=0]{fig8b.epsi}
\includegraphics[width=60mm,angle=0]{fig8c.epsi}
\hspace{0.5cm}
\includegraphics[width=60mm,angle=0]{fig8d.epsi}
\includegraphics[width=60mm,angle=0]{fig8e.epsi}
\hspace{0.5cm}
\includegraphics[width=60mm,angle=0]{fig8f.epsi}
\caption{Properties of the MAGMA clouds, plotted as a function of the
  interstellar pressure, $P_{h}$. The panels, plot symbols and
  annotations are the same as in Fig.~\ref{fig:isrfcmp}.}
\label{fig:phcmp}
\end{center}
\end{figure*}

\noindent The comparison between properties of the MAGMA clouds and
$\Sigma_{g}$ is helpful for interpreting the plots in
Fig.~\ref{fig:phcmp}. In particular, both $\sigma_{\rm v}$ and
$\Sigma_{\rm H_{2}}$ increase in regions with higher $P_{h}$,
following their behaviour in regions with high \hi\ column
density. The dominant physics underlying these correlations may
therefore be the importance of an atomic shielding layer for the
survival of \hh\ molecules, rather than pressure
regulation. $\Sigma_{\rm H_{2}}$ and $P_{h}$ is more strongly
correlated with $P_{h}$ than with $\Sigma_{g}$ however, and we observe
that there are few clouds with low $I_{\rm CO}$ or $\Sigma_{\rm
  H_{2}}$ in regions with high $\Sigma_{*}$ (panels [d] and [e] of
Fig.~\ref{fig:stellarcmp}). Insofar as $I_{\rm CO}$ and $\Sigma_{\rm
  H_{2}}$ are reliable tracers of the \hh\ surface density, this
provides some indication that shielding alone may not regulate the
\hh\ surface density in regions of the LMC where the interstellar gas
pressure is high.\\

\section{Discussion}
\label{sect:discussion}

\subsection{Properties of young GMCs}
\label{sect:fukuigmcmodel}

\noindent In Section~\ref{sect:younggmcs}, we investigated the
physical properties of the young GMC candidates in the MAGMA LMC cloud
list, finding that sightlines through these clouds tend to have lower
$T_{\rm max}$ and $\langle T_{\rm pk} \rangle$ than sightlines through
star-forming GMCs of equivalent size. Models of the \aco\ emission
from GMCs indicate that the emission arises from a large number of
optically thick clumps that are not self-shadowing
\citep[e.g.][]{wolfireetal93}. For each independent sightline through
a GMC, the observed peak CO brightness is then a measure of the total
projected area of the optically thick clumps within the beam area,
weighted by their brightness temperature, which should correspond to
the kinetic temperature of the CO-emitting gas if emission in the
clumps is optically thick \citep[e.g][]{maloneyblack88}. The usual
assumption is that the distribution of clump sizes and the brightness
temperature do not vary significantly between sightlines, and that the
observed CO brightness measures the number of clumps -- and hence the
total amount of molecular gas -- within the telescope beam. While this
assumption may be justified for clouds in the inner disc of the Milky
Way (S87), it is worth noting that the peak CO brightness results from
a combination of the brightness temperature, the number of CO-emitting
clumps within the beam, and their average size.  A possible
interpretation for the different average peak CO brightness of
star-forming and non-star-forming GMCs is that there are fewer
CO-emitting clumps in GMCs without star formation, leading to a lower
angular filling factor of CO emission. Alternatively, colder gas
temperatures in the dormant CO-emitting substructures within
non-star-forming GMCs could lead to a lower average brightness
temperature for the clump ensemble. \\

\noindent As our young GMC candidates tend to have a lower $\langle
T_{\rm pk} \rangle$ and $T_{\rm max}$ values than star-forming GMCs of
comparable size, it might be expected that their total CO luminosity
would also be fainter. However, the KS tests reveal no systematic
differences between the distributions of $L_{\rm CO}$ and $I_{\rm CO}$
for the various cloud samples, which suggests that the regions of high
CO brightness in the star-forming GMCs are restricted to a relatively
small number of pixels in the MAGMA data subcubes. This is consistent
with the view of star formation as a highly localised process:
occupying only a small fraction of the total cloud volume,
star-forming clumps have temperatures and densities that are much
higher than in the bulk of the GMC, most of which does not participate
directly in star formation. The lower $\langle T_{\rm pk} \rangle$ and
$T_{\rm max}$ of non-star-forming GMCs would seem to provide some
preliminary evidence that the general characteristics of substructure
within non-star-forming GMCs in the Milky Way -- i.e. cooler clumps
that are less massive and more diffuse than in star-forming GMCs
\citep[e.g.][]{williamsetal94, williamsblitz98} -- will also be found
to apply in the LMC. \\

\begin{figure*}
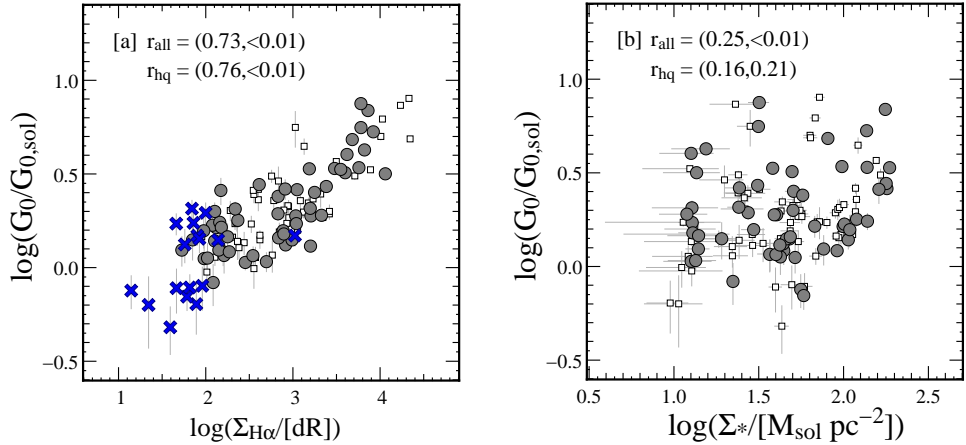

\begin{center}
\includegraphics[width=60mm]{fig9a.epsi}
\hspace{0.5cm}
\includegraphics[width=60mm]{fig9b.epsi}
\caption{The radiation field at the location of the MAGMA clouds,
  plotted as a function of (a) the \ha\ surface brightness, as
  observed by SHASSA \citep{gaustadetal01}, and (b) the stellar mass
  surface density. Plot symbols are the same as in
  Fig.~\ref{fig:Rdv}. The Spearman correlation coefficient and
  corresponding $p$-value for all 125 MAGMA GMCs and the high quality
  subsample are indicated at the top left of each panel.}
\label{fig:isrf_dist}
\end{center}
\end{figure*}

\noindent Finally, we found a clear trend for the young GMC candidates
to be located in regions where $G_{0}$ is relatively weak. A
straightforward explanation for this result is that once young massive
stars begin to form, they are an important source of dust-heating
within LMC molecular clouds. Importantly, this would mean that $G_{0}$
is not strictly tracing the ambient (i.e. external) radiation field
incident on a GMC, but instead contains a signficant contribution from
the massive stars that are within -- or have recently emerged from --
their GMC progenitor. A strong correlation between $G_{0}$ and
\ha\ surface brightness for all the MAGMA LMC clouds (see
Fig.~\ref{fig:isrf_dist}) indicates a strong connection between dust
heating and star-formation activity, which would seem to support this
interpretation.\\

\subsection{The origin of Larson's laws}
\label{sect:originll}

\noindent Despite the longevity of Larson's scaling relations, a
complete theoretical explanation for the origin of the size-linewidth
relation is still lacking. Considerable effort has been devoted to
demonstrating that the observed size-linewith relationship can be
reproduced by realistic models of interstellar turbulence
\citep[e.g.][and references
  therein]{maclowklessen04,ballesterosparedesetal07}, and that
turbulence in the cold gas phase is both universal and nearly
invariant \citep[e.g.][]{heyerbrunt04}, but differences between the
working variables and descriptive tools available to theory and
observation represent a significant obstacle to making turbulent
models empirically falsifiable. In many instances, however, older
explanations for the origin of Larson's laws have not been thoroughly
tested, since there are few extragalactic datasets that are comparable
to the Milky Way molecular cloud samples that guided the development
of these early models. In this section, we therefore consider whether
the MAGMA data are consistent with two analytic models for the origin
of Larson's laws: the photoionization-regulated star formation theory
proposed by M89 and the model of molecular clouds as virialized
polytropes proposed by E89. 

\subsubsection{Comparison with \citet{mckee89}}
\label{sect:mckee89}

\noindent The photoionization-regulated theory of star formation
proposed by M89 provides one possible explanation for the origin of
Larson's laws. In this theory, molecular clouds evolve towards an
equilibrium state where energy injection from newborn low-mass stars
halts the clouds' gravitational contraction. Equilibrium depends on
the level of photoionization by the interstellar far-ultraviolet (FUV)
radiation field because ambipolar diffusion governs the rate of
low-mass star formation in the cloud; in turn, the rate of ambipolar
diffusion is regulated by the ionization fraction, which depends on
the interstellar FUV radiation field in the bulk of the molecular
cloud. A second factor that determines the equilibrium state of the
GMCs is the local dust abundance, since dust shields \hh\ molecules
against photodissociation. Notably, the M89 theory predicts that
molecular clouds in equilibrium should have uniform extinction, rather
than constant column density: clouds in environments with low
dust-to-gas ratios and/or strong radiation fields will require larger
column densities to attain the equilibrium level of extinction, and
should therefore follow a $R-\sigma_{\rm v}$ relation with a higher
coefficient than the Milky Way relation. Recent measurements of the
surface densities of extragalactic GMCs contradict this prediction,
instead showing that molecular cloud surface densities in low
metallicity environments are similar to, or even lower than, the
average mass surface density of Milky Way clouds
\citep[e.g. B08,][]{leroyetal07}. The median mass surface density of
the MAGMA clouds is $\sim50$~\mpcsq, in line with these results.\\

\noindent It is worth noting, however, that low mass surface densities
(i.e. less than $\sim 100$~\mpcsq\ for GMCs in sub-solar metallicity
environments) are not necessarily inconsistent with the M89 model: the
complete prediction by M89 is that the velocity dispersion of
molecular clouds should increase as the dust-to-gas ratio decreases,
provided that the densities of the CO-emitting clumps within GMCs are
comparable to the densities of the clumps in Galactic molecular
clouds, and that appropriate corrections for the angular filling
fraction of the CO-emitting clumps, $\mathcal{C}$, have been
applied. If the CO-emitting clumps within GMCs have $\mathcal{C}
\sim1$, densities of $n_{\rm H} = 10^{3}$~\ccc, and clump-to-cloud
extinction ratios of $\bar{A_{c}}/\bar{A_{v}} \sim 0.3$ -- values that
M89 regards as typical for molecular clouds in the inner Milky Way
disc -- then the photoionization-regulated star formation theory
predicts that equilibrium is achieved for visual extinctions between
$\bar{A_{v}}\sim4$ and 8~mag. Equations 5.5 to 5.7 in M89 show that an
equilibrium extinction value of $\bar{A_{v}} \sim 1$ can none the less
be obtained in an environment with the LMC's typical dust-to-gas ratio
\citep[$\sim 0.3$, ][]{dobashietal08} if the CO-emitting clumps within
GMCs have an angular filling fraction of $\mathcal{C} = 0.25$ and a
typical density of $n_{\rm H} = 10^{4.5}$~\ccc, and if the fraction of
the total cloud mass residing in these dense clumps is $\sim45$ per
cent. While agreement with the M89 model is theoretically possible, it
requires the density contrast in GMCs in the LMC to be more extreme
than in Milky Way clouds. $\mathcal{C}$ values less than unity have
previously been invoked to explain the low $T_{\rm pk}$ measurements
for LMC clouds \citep[e.g.][]{wolfireetal93,kutneretal97,garayetal02},
but average densities of $n_{\rm H} \sim 10^{4.5}$~\ccc\ for the
CO-emitting clumps seem less plausible. Excitation analyses of
millimetre and submillimetre spectral line observations
\citep[e.g.][]{heikkilaetal99,pinedaetal08trunc,minamidanietal08trunc}
have reported clump densities between $10^{4}$ and $10^{6}$~\ccc, but
as these studies targeted the LMC's brightest star-forming regions it
remains uncertain whether similar clump densities would be common in
molecular clouds throughout the LMC. \\

\noindent A further problem for the photoionization-regulated star
formation theory is that there is no sign of a correlation between
$G_{0}$ and $\Sigma_{\rm H_{2}}$ or $I_{\rm CO}$ for GMCs without
signs of star formation (blue crosses in panels [d] and [e] of
Fig.~\ref{fig:isrfcmp}). In the scenario postulated by M89,
equilibrium is only achieved prior to the onset of massive star
formation: after this, young massive stars rapidly disrupt their natal
clouds, destroying the relationship between the gas column density,
dust abundance and ambient radiation field. The GMCs that are
designated as star-forming in the MAGMA sample contain at least one O
star \citep{kawamuraetal09} and their young stellar content makes a
significant contribution to the radiation field within the cloud (see
Table~\ref{tbl:ks1}). For GMCs without star formation, however,
$\Sigma_{\rm H_{2}}$ and $I_{\rm CO}$ should show a correlation with
$G_{0}$ if their column density is regulated by the ambient radiation
field, assuming that the dust-to-gas ratio is roughly constant in the
environments of non-star-forming GMCs across the LMC. No such
correlation is apparent in panel [d] or [e] of Fig.~\ref{fig:isrfcmp}.\\

\noindent Finally, we note that a correlation between the
internally-generated radiation field and the \hh\ mass surface density
would be expected if the star formation efficiency of molecular gas
were constant \citep[][]{leroyetal08}, since GMCs with higher
\hh\ column densities -- and presumably higher volume densities --
should have a higher surface density of star formation. The
correlation tests in Section~\ref{sect:tracers} indicated that
$\Sigma_{H_{2}}$ for star-forming GMCs in the high quality subsample
are associated with higher values of $G_{0}$, but the trend is not
significant if we consider all the GMCs in the MAGMA cloud list, or if
we use $I_{\rm CO}$ rather than $\Sigma_{\rm H_{2}}$ to trace the
\hh\ mass surface density.  \\

\subsubsection{Comparison with \citet{elmegreen89}}
\label{sect:e89}

\noindent Another explanation for the origin of Larson's laws was put
forward by E89, who proposed that molecular clouds and their atomic
envelopes are virialized, magnetic polytropes (i.e. with internal
pressure $P$ that varies with the density $\rho$ according to $P = K
\rho^{n}$, where $K$ is a constant and $n$ is the polytropic index)
and with external pressure that is determined by the kinetic pressure
of the interstellar medium. In this theory, the radius and density of
a molecular cloud complex adjust so that the pressure at the boundary
of the atomic envelope equals the ambient kinetic pressure. In this
case, molecular clouds within a galaxy have a similar mass surface
density because the ambient pressure throughout the galaxy is also
roughly uniform. While the notion that molecular clouds can achieve
dynamical equilibrium within their lifetime has been disputed
\citep[e.g.][]{hartmannetal01}, recent work has highlighted the
potential importance of pressure for the formation of molecular gas
\citep[e.g.][]{wongblitz02,blitzrosolowsky04,blitzrosolowsky06}.\\

\noindent The molecular cloud model proposed by E89 predicts that
$C_{0}$ and $\Sigma_{\rm H_{2}}$ should scale with the total mass of
the atomic+molecular cloud complex, due to the increasing weight of
the atomic gas layer surrounding large molecular clouds. In
Sections~\ref{sect:Rdv} and~\ref{sect:tracers}, we saw that the
scatter in the Larson-type scaling relations for the MAGMA clouds
implies order of magnitude variations in $C_{0}$ and $\Sigma_{\rm
  H_{2}}$, and also evidence for a weak correlation between
$\Sigma_{\rm H_{2}}$ and the interstellar pressure
(Fig.~\ref{fig:phcmp}[e]). These trends motivate us to examine more
closely whether the MAGMA clouds are consistent with the E89 theory
for the origin of Larson's Laws, once variations in the local
radiation field, metallicity and mass of atomic gas surrounding the
GMCs are taken into account. \\

\noindent In lieu of Larson's third law, E89 predicts that the total
(i.e. atomic+molecular) mass surface density of molecular cloud
complexes will vary according to
\begin{equation}
\frac{M_{t}}{R_{t}^{2}} \approx 190\pm90 \left(\frac{P_{e}}{\phu}\right)^{1/2} \mpcsq,
\label{eqn:e89eqn12}
\end{equation}
where $P_{e}$ is the external pressure on a molecular+atomic cloud
complex, $M_{t}$ is the total mass of the atomic+molecular complex and
$R_{t}$ is the radius of the complex. This equation can be written in
terms of observed molecular cloud properties:
\begin{equation}
\frac{M_{m}}{R_{m}^{2}} \approx 380\pm250 \left(\frac{P_{e}}{\phu}\right)^{-1/24} \left(\frac{M_{t}}{10^{5} \msol}\right)^{1/4} \left(\frac{G_{0}/G_{\odot,0}}{Z/Z_{\odot}}\right)^{1/2} \mpcsq,
\label{eqn:e89eqn17}
\end{equation}
where $R_{m}$ and $M_{m}$ are the radius and mass of the molecular
part of the cloud complex. The range of coefficients in these
equations arise from solutions to the virial theorem for plausible
values of the adiabatic index and variations in the ratio of the
magnetic field pressure to the kinetic pressure (see fig.~3 in
E89). To explore the relative importance of the parameters that
contribute to the right hand side of Equation~\ref{eqn:e89eqn17}, we
defined six basic models that use different estimates for the total
mass, metallicity and external radiation field of the molecular cloud
complexes. For all models, $P_{e}$ is assumed to be the same as the
kinetic pressure of the ambient interstellar medium, i.e. $P_{e} =
P_{h}/(1+\alpha+\beta)$ where $\alpha\sim0.4$ and $\beta\sim0.25$ are
the relative contribution by cosmic-rays and the magnetic field to the
total pressure, and $P_{h}$ is estimated according to
Equation~\ref{eqn:pressure}. The model parameters are summarized in
Table~\ref{tbl:e89mdls}.\\

\begin{table}
\caption{A summary of the model parameters which are substituted into
  Equation~\ref{eqn:e89eqn17} in order to produce the plots in
  Fig.~\ref{fig:e89ll}. The first column describes the metallicity:
  either a shallow radial gradient of $-0.05$~dex~kpc$^{-1}$
  \citep[e.g. ][]{coleetal04}, or a constant $Z/Z_{\odot} = 0.5$ for
  all clouds in assumed. We adopt 05h19m30s -68d53m (J2000) for the
  location of the LMC's kinematic centre, and $i=35^{\circ}$ for the
  LMC's inclination \citep{wongetal09}. The second column lists $k$,
  the factor by which the measured value of $G_{0}/G_{0,\odot}$ has
  been reduced for star-forming clouds in the MAGMA cloud list. We
  introduce $k$ to account for the fact the radiation field in
  Equation~\ref{eqn:e89eqn17} is the external field incident on the
  molecular cloud, i.e. it does not include the contribution from
  young massive stars within the cloud. The third column describes the
  relationship between the total atomic+molecular mass, $M_{t}$, and
  the molecular mass, $M_{m} \equiv M_{\rm vir}$, that we adopt for
  the clouds. For M1 to M4, we use the mass dependence adopted by
  E89. The mass dependencies for M5 and M6 are constructed such that
  the total mass of the cloud complex increases more rapidly than the
  molecular mass; the coefficient of the mass dependence for these
  models is then chosen such that $M_{t} \geq M_{m}$ for the observed
  range of MAGMA GMC masses.}
\label{tbl:e89mdls}
\begin{tabular}{@{}cccc}
\hline
Model Identifier & $Z/Z_{\odot}$ & $k$ & Mass Dependency \\
\hline
M1 & 0.5 & 1.0 & $M_{t} = 2 M_{m}$ \\ 
M2 & 0.5 & 2.0 & $M_{t} = 2 M_{m}$ \\ 
M3 & Gradient & 1.0 & $M_{t} = 2 M_{m}$ \\ 
M4 & Gradient & 2.0 & $M_{t} = 2 M_{m}$ \\ 
M5 & Gradient & 1.0 & $M_{t}/[\msol] = (M_{m}/21.45~\msol)^{1.5}$ \\ 
M6 & Gradient & 1.0 & $M_{t}/[\msol] = (M_{m}/141.42~\msol)^{2.0}$  \\
\hline
\end{tabular}
\end{table}

\begin{figure*}
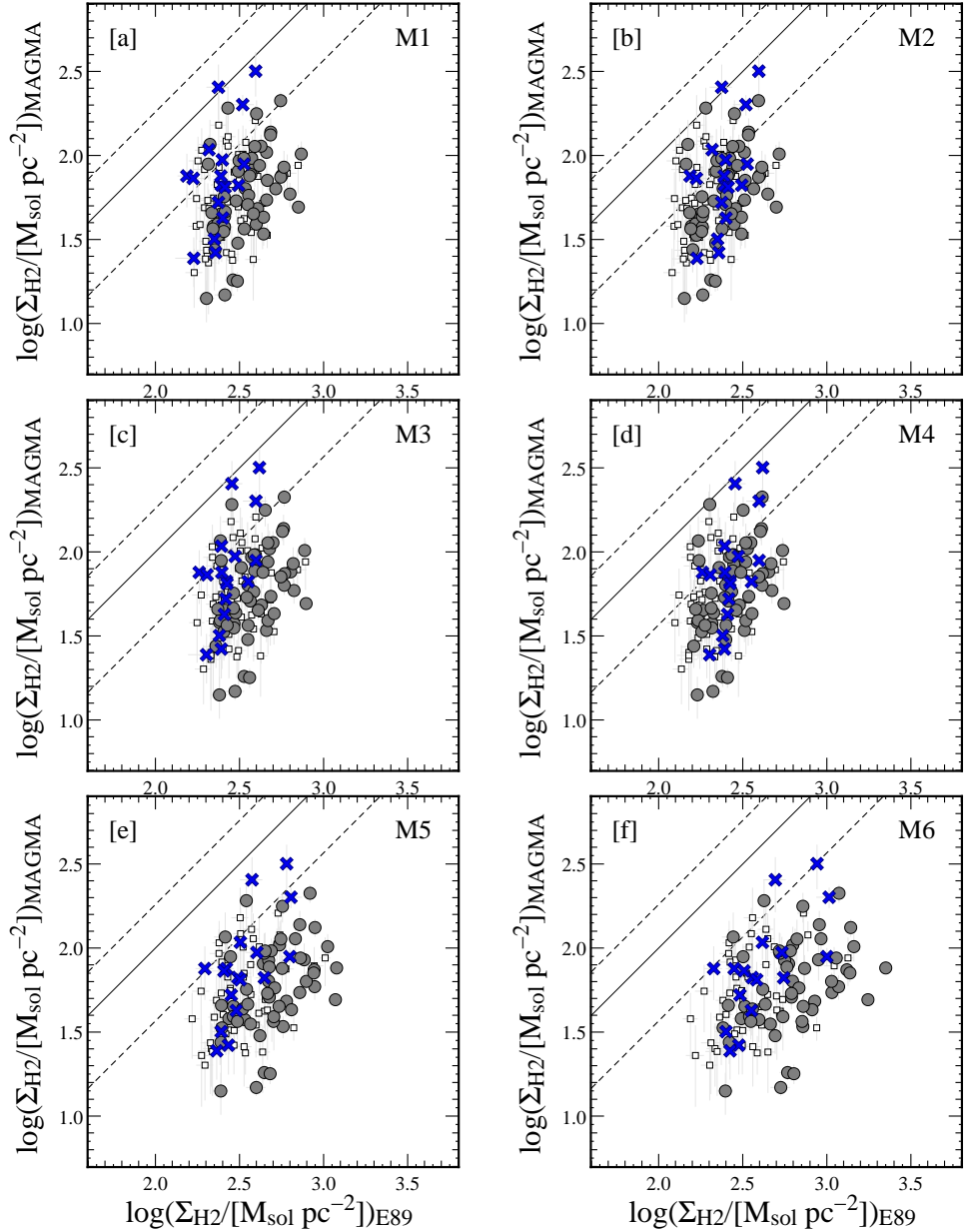

\begin{center}
\includegraphics[width=60mm,angle=0]{fig10a.epsi}
\hspace{0.5cm}
\includegraphics[width=60mm,angle=0]{fig10b.epsi}
\includegraphics[width=60mm,angle=0]{fig10c.epsi}
\hspace{0.5cm}
\includegraphics[width=60mm,angle=0]{fig10d.epsi}
\includegraphics[width=60mm,angle=0]{fig10e.epsi}
\hspace{0.5cm}
\includegraphics[width=60mm,angle=0]{fig10f.epsi}
\caption{The observed values of $\Sigma_{\rm H_{2}}$ for the
  MAGMA clouds versus the values predicted by E89 for the models
  of the metallicity, external radiation field and total cloud mass
  presented in Table~\ref{tbl:e89mdls}. Plot symbols are the same as in
  Fig.~\ref{fig:Rdv}. The solid line represents equality
  between the observed and predicted values for the coefficients
  adopted by E89. The dashed lines indicate the potential shift in the
  line of equality for the maximum range of the coefficients in
  Equation~\ref{eqn:e89eqn17}.}
\label{fig:e89ll}
\end{center}
\end{figure*}

\noindent In Fig.~\ref{fig:e89ll}, we plot the observed values of
$\Sigma_{\rm H_{2}}$ for the MAGMA clouds against the values predicted
by Equation~\ref{eqn:e89eqn17} for each of the six
models. Fig.~\ref{fig:e89ll} shows that the measured values for
$\Sigma_{\rm H_{2}}$ for the MAGMA clouds are significantly lower than
predicted by E89, regardless of which model we adopt. The average mass
surface density is sensitive to variations in the metallicity and
radiation field, but the LMC would need to have $Z \geq Z_{\odot}$ and
$G_{0} \leq G_{0,\odot}$ for the MAGMA clouds to be consistent with
the closest line of equality in Fig.~\ref{fig:e89ll} (which
corresponds to the lowest value of the coefficient in
Equation~\ref{eqn:e89eqn17}). The models in which the total complex
mass increases faster than the molecular mass (M5 and M6) demonstrate
a better agreement with the slopes predicted by E89, but these models
are not physically realistic since they imply that the mass of the
\hi\ envelopes surrounding the MAGMA clouds is almost an order of
magnitude greater than the total \hi\ mass of the LMC \citep[$5 \times
  10^{8}~\msol$, ][]{staveleysmithetal03}. Reducing the coefficient of
the mass dependency in these models would lower the total
\hi\ envelope mass and also shift the data towards the line of
equality in Fig.~\ref{eqn:e89eqn17}, but it leads to the unphysical
solution that $M_{t} < M_{m}$ for low mass MAGMA clouds.\\

\noindent The total atomic+molecular mass of the cloud complex is
perhaps the most important source of uncertainty in the present
comparison, and we are currently undertaking an analysis of the
\hi\ and MAGMA CO datasets that should place empirical constraints on
the mass of the \hi\ gas that is associated with individual MAGMA
clouds. However, two further comments regarding the \hi\ data and its
potential to constrain the E89 model are worth noting here. First,
visual inspection of the CO and \hi\ LMC maps shows that the majority
of GMCs are not isolated, but are instead located within a group of
molecular clouds that appear to share a common \hi\ envelope. This
possibility is not explicitly addressed by E89, but it is important
insofar as neighbouring molecular clouds will partially shield each
other, reducing the total amount of \hi\ required for the shielding
layer. Second, we note that CO emission is detected in regions where
the {\it total} \hi\ column density through the LMC is only $\nh
\approx 1 - 2 \times 10^{21}$~\pcmsq\ \citep[refer fig. 2
  of][]{wongetal09}. This is problematic insofar as the
\hh\ self-shielding theory invoked by E89 requires a uniform shielding
layer with $\nh = 1.2 \times 10^{21}$~\pcmsq\ for a molecular cloud
situated in an ambient radiation field $G_{0}/ G_{0,\odot} = 1.5$ and
metallicity $Z/Z_{\odot} = 0.5$, assuming an density of $n \approx
60$~H~\ccc\ in the atomic layer \citep{federmanetal79}. $\nh = 1.2
\times 10^{21}$~\pcmsq\ is probably a lower limit: if the gas in the
shielding layer is clumpy, then the required column density increases
by a factor $f/f_{\odot}$, where $f = \langle n \rangle^{2}/\langle
n^{2} \rangle$ is the volume filling factor. Part of this discrepancy
may be because sightlines towards GMCs in the LMC have significant
optical depth: \hi\ absorption studies indicate peak optical depths
between $\tau = 0.4$ and 2.0 near regions with CO emission, implying
that the true \hi\ column density is a factor of $\tau/[1 - e(-\tau)]
= 1.2 - 2.3$ greater than the observed value
\citep{dickeyetal94,marxzimmeretal00}. A recent study of the LMC's
far-infrared emission has also proposed a widespread cold atomic gas
component with significant optical depth in the LMC in order to
explain an excess of emission at 70~$\mu$m
\citep{bernardetal08trunc}. \\

\noindent More generally, we note that plausible variations in the
interstellar pressure due to e.g. a widespread ionized gas component
or a high rate of cosmic-ray escape would not have a significant
impact on the plots in Fig.~\ref{fig:e89ll}, since the exponent of
the pressure term in Equation~\ref{eqn:e89eqn17} is small. A
significant mass of \hh\ without CO emission associated with each GMC
would tend to rather magnify than reduce the discrepancy between the
MAGMA clouds and the E89 model, moreover, except in a physically
improbable scenario where the CO-dark \hh\ gas has a greater average
density than the CO-emitting region of the GMC. One possible
resolution may be the porosity of the LMC's interstellar medium: as
noted above, the calculations in E89 assume a uniform gas layer which
provides more effective shielding than gas that is highly clumped. A
comparison between the volume filling factor of \hi\ in the solar
neighbourhood and in the LMC, plus the inclusion of clumpy cloud
structure in the E89 theory, would be required to test this hypothesis
however.

\subsection {GMC properties: trends with environment}
\label{sect:envirotrends}

\noindent Previous comparative studies of the Milky Way and nearby
extragalactic GMC populations have found that GMC properties are
relatively uniform, and mostly insensitive to variations in
environmental parameters such as the radiation field and dust
abundance \citep[e.g. B08, ][]{rosolowskyetal03,rosolowsky07}. A
possible explanation for this result is that GMCs are strongly bound
and hence largely decoupled from conditions in the local ISM. B08 have
cautioned, however, that the apparent universality of GMC properties
as measured from CO observations may simply reflect the physical
conditions required for CO emission to be excited, and that the
properties and extent of the \hh\ material surrounding a GMC's high
density CO-emitting peaks might indeed be sensitive to local
environmental factors, a conclusion that is supported by far-infrared
studies of molecular clouds in nearby dwarf galaxies
\citep[e.g.][]{israel97,leroyetal07,leroyetal09}. Our analysis tends
to support the view that CO-derived properties of GMCs are mostly
insensitive to environmental conditions, but there are several
exceptions that we discuss below.

\subsubsection{GMC properties and $\Sigma_{*}$}

\noindent In Section~\ref{sect:cmpstellar}, we showed that $\langle
T_{\rm pk} \rangle$ and $I_{\rm CO}$ increase for GMCs located in
regions of high $\Sigma_{*}$, without a corresponding variation in $R$
or $\sigma_{\rm v}$. A key variable to explain this result may be the
role of the old stars in dust production, as a recent analysis of {\it
  Spitzer} mid- and far-infrared data for the LMC indicates that VSGs
and polycyclic aromatic hydrocarbon (PAH) molecules are overabundant
in the LMC's stellar bar \citep{paradisetal09}. More specifically, if
the abundance of VSGs and/or PAHs were increased by the ejecta from
mass-losing old stars, then photoelectric heating of the molecular gas
may be more efficient in regions with high $\Sigma_{*}$, potentially
raising the CO excitation temperature. Alternatively, a higher overall
dust abundance could reduce the selective photodissociation of CO
molecules and lead to a higher abundance of CO relative to \hh\ for
GMCs in the stellar bar. This could occur if \hh\ readily
self-shields, while the survival of CO molecules relies more on the
attenuation of the photodissociating radiation by dust
\citep[e.g.][]{maloneyblack88}. Although the CO emission from Milky
Way molecular clouds is approximately independent of variations in the
CO abundance -- firstly due to the high optical depth of the
\aco\ line, and secondly because the angular filling factor of the
CO-emitting clumps at any particular radial velocity within the cloud
is $\sim1$ \citep[see e.g.][]{wolfireetal93} -- this may not be true
of LMC molecular clouds \citep[see e.g.][]{maloneyblack88}. The low
values of $T_{\rm pk}$ for \aco\ emission in LMC clouds suggests that
the angular filling factor of the CO-emitting clumps is indeed
relatively small \citep[see
  e.g.][]{kutneretal97,johanssonetal98,garayetal02}, in which case
higher CO-to-\hh\ ratios might produce higher values of $\langle
T_{\rm pk} \rangle$ and $I_{\rm CO}$.\\

\noindent Contrary to what might be expected if higher interstellar
pressures promote the formation of molecular gas, there is no evidence
that $R$ or $\sigma_{\rm v}$ for the LMC molecular clouds increases in
regions with high $\Sigma_{*}$. The absence of these correlations is
perhaps not very significant, however, since our estimate for $P_{h}$
is dominated by $\Sigma_{g}$ for regions with $\Sigma_{*} \leq
60$~\mpcsq; if the \hi\ emission along sightlines towards GMCs has
significant optical depth, $\Sigma_{g}$ will dominate $P_{h}$ to even
higher $\Sigma_{*}$ thresholds ($\Sigma_{*} \sim 100$~\mpcsq\ for
$\tau = 1$). A correlation between $G_{0}$ and $\Sigma_{*}$ might be
expected if regions with strong stellar gravity promoted the formation
of overdensities within GMCs and enhanced the local star formation
rate \citep[e.g.][]{leroyetal08} or, alternatively, if old stars made
a significant contribution to the dust heating. Stronger external
heating in the stellar bar region has previously been invoked as an
explanation for the higher $\bco/(J=1\to0)$ intensity ratios in
inner LMC molecular clouds \citep{soraietal01}. While the correlation
tests in Section~\ref{sect:tracers} provided no clear evidence for a
relationship between $G_{0}$ and $\Sigma_{*}$, we do not observe
clouds with low $G_{0}$ values located in regions with $\Sigma_{*}
\geq 100$ ~\mpcsq\ (see Fig.~\ref{fig:isrf_dist}[b]), so it is possible
that enhanced star formation due to strong stellar gravity and/or dust
heating by old stars becomes important at higher stellar densities. 

\subsubsection{GMC properties and \nh}

\noindent In Section~\ref{sect:cmphi}, we found a weak but significant
correlation between \nh\ and $\sigma_{\rm v}$
(Fig.~\ref{fig:nhcmp}[b]). Its interpretation is somewhat uncertain,
however, since there is no obvious trend between \nh\ and the GMC
radius, which would be expected if larger \hi\ column densities were
associated with more massive GMCs (assuming that the average density
of GMCs also remains constant across the LMC). Some positive
association between \nh\ and $R$ would be expected, moreover, simply
as a consequence of the size-linewidth relation for LMC molecular
clouds. Perhaps the simplest explanation is that the mass and size of
GMCs increase with \nh, but that the GMC radius derived from CO
observations is not a sensitive tracer of the true cloud size. Another
possibility is that the density of molecular clouds genuinely
increases with the local \hi\ column density; this could account for
some of the scatter in the $R-\sigma_{\rm v}$ relation but would imply
that $\Sigma_{\rm H_{2}}$ and \xco\ -- assuming the average CO
brightness temperature is constant -- should also increase with
\nh\ \citep[see e.g.][]{dickmanetal86,heyeretal01}.  A third
possibility is that the degree of virialisation in GMCs varies across
the LMC, i.e. GMCs in regions with high \nh\ are less gravitationally
bound than GMCs in regions with low \nh, due to the higher external
pressure. Provided that the average CO brightness temperature of GMCs
in the LMC remains constant, we would also expect the observed value
of \xco\ to increase with \nh\ in this case.  Our analysis in
Section~\ref{sect:tracers} revealed some evidence for trends between
$\Sigma_{\rm H_{2}}$, \xco\ and \nh, although the correlations
involving the high quality subsample did not satisfy our criteria for
significance. There is sufficient scatter in Figures~\ref{fig:Rdv}
and~\ref{fig:nhcmp} that none of these explanations can be
definitively ruled out. 

\subsubsection{GMC properties and $P_{h}$}

\noindent Although the MAGMA clouds do not follow the predictions of
the E89 molecular cloud model, we find that $\sigma_{\rm v}$ and
$\Sigma_{\rm H_{2}}$ increase with the interstellar pressure (panels
[b] and [e] of Fig.~\ref{fig:phcmp}). These correlations suggest that
the external pressure on a GMC in the LMC may indeed play a key role
in regulating its dynamical properties, although we caution that we
cannot readily distinguish between the role of pressure and shielding
in our analysis due to the dominant contribution of $\Sigma_{g}$ to
our estimate for $P_{h}$. The low absolute values of $\Sigma_{\rm
  H_{2}}$ and $I_{\rm CO}$ for the MAGMA clouds imply that their
average internal pressure is also low. From the virial theorem, a GMC
with $\Sigma_{\rm H_{2}} \sim 50$~\mpcsq\ will have internal pressure
$P_{h}/k_{B} \sim 5 \times
10^{4}$~K~\ccc\ \citep[e.g. ][]{krumholzmckee05}; this is not much
greater than the average external kinetic pressure for the MAGMA GMCs,
$\langle P_{h}/k_{B} \rangle \sim 3.9 \times 10^{4}$~K~\ccc, estimated
from Equation~\ref{eqn:pressure}. Plausible optical depth corrections
for the \hi\ emission along sightlines towards GMCs (i.e. for $\tau$
values between 0.4 and 2) would effectively balance these estimates
for the average pressure internal and external to the cloud, although
the dense clumpy structure within the GMC would be likely to remain
significantly overpressured. A gentle pressure gradient across the
molecular cloud boundary may explain why we have found that some
properties of the MAGMA clouds are correlated with environmental
conditions. It would be interesting to test whether GMCs in nearby
spiral galaxies -- which have $\langle \Sigma_{\rm H_{2}} \rangle \sim
130$~\mpcsq\ (B08) and presumably higher internal pressures relative
to the surrounding ISM \citep[see also ][]{krumholzetal09b} -- exhibit
any of the correlations that we have identified for the MAGMA GMCs.

\section{Conclusions}
\label{sect:conclusions}

\noindent This paper has presented Mopra Telescope observations of
\aco\ emission from a sample of 125 GMCs in the Large Magellanic
Cloud. The data described here were obtained as part of MAGMA, an
ongoing mapping survey of the \aco\ emission from molecular gas in the
Magellanic Clouds. The MAGMA data, which have an angular resolution of
45 arcsec and a spectral resolution of 0.1\kms, will be made
available to the astronomical community upon completion of the survey,
and should be a rich resource for follow-up studies with millimetre
and submillimetre facilities on the Atacama Plateau. A discussion of
MAGMA's observational strategy, data products and a final GMC
catalogue will be presented in a future paper (Wong \ea, in
preparation). In this paper, we examined the empirical scaling
relations between the basic physical properties of the GMCs in our
current LMC cloud list, and dependencies between the cloud properties
and the local interstellar environment. We report the following
results and conclusions: \\

\noindent 1. The observed GMCs have radii ranging between 13 and
160~pc, velocity dispersions between 1.0 and 6.1~\kms, peak CO
brightnesses between 1.2 and 7.1~K, CO luminosities between $10^{3.5}$
and $10^{5.5}$~K~\kms~pc$^{2}$, and virial masses between $10^{4.2}$
and $10^{6.8}$~\msol. The clouds tend to be elongated, with a median
major-to-minor axis ratio of 1.7. These values are comparable to the
measured properties of Galactic GMCs. The average mass surface density
of the observed clouds is $\sim50$~\msol~pc$^{-2}$, approximately half
the value determined for GMCs in the inner Milky Way catalogue of
\citet{solomonetal87}. \\

\noindent 2. The MAGMA clouds exhibit scaling relations that are
similar to those previously determined for Galactic and extragalactic
GMC samples \citep[e.g.][]{solomonetal87,bolattoetal08}. However the
MAGMA LMC clouds are offset towards narrower linewidths and lower CO
luminosities compared to GMCs of a similar size in these samples. The
scatter in the scaling relations corresponds to order of magnitude
peak-to-peak variations in the CO-to-\hh\ conversion factor (as
inferred from the ratio of the virial mass to the CO luminosity), the
\hh\ mass surface density and the CO surface brightness of the MAGMA
GMCs.\\

\noindent 3. The physical properties of star-forming GMCs are very
similar to the properties of GMCs without signs of massive star
formation. Sightlines through non-star-forming GMCs tend to have lower
peak CO brightness, suggesting that the filling fraction and/or
brightness temperature of the CO-emitting substructure is lower for
clouds without star formation.   \\

\noindent 4. We find a significant positive correlation between the
peak CO brightness and CO surface brightness of the MAGMA clouds and
the stellar mass surface density. We propose that these correlations
are due to an increase in the CO brightness temperature and/or an
increase in the abundance of CO relative to \hh\ in the stellar bar
region. \\

\noindent 5. The velocity dispersion ($\sigma_{\rm v}$) of the MAGMA
GMCs increases in regions with high \hi\ column density (\nh). Higher
volume densities and/or higher virial parameters for GMCs in regions
with high \nh\ could produce the observed correlation, although the
MAGMA data does not provide unambiguous evidence for either of these
alternatives.\\

\noindent 6. There is some evidence that the \hh\ mass surface density
of the MAGMA LMC clouds increases with the interstellar kinetic
pressure, $P_{ext}$. Although the molecular cloud model proposed by
\citep{elmegreen89} predicts a relation between $P_{ext}$ and the mass
surface density of an atomic+molecular cloud complex, the MAGMA clouds
do not fulfil the predictions of the model for reasonable values of
the metallicity, radiation field and GMC envelope mass. 


\section*{Acknowledgments}
\label{sect:thanks}
\noindent We thank Christian Henkel, Jonathan Seale and Min Wang for
their assistance with MAGMA observations. We also thank staff at the
ATNF for observing support, and Robert Gruendl for providing the
stellar mass surface density image. We acknowledge extensive use of
NASA's Astrophysics Data System Bibliographic Services. TW is
supported by NSF grant 08-07323 and the University of Illinois. JO is
supported by the National Radio Astronomy Observatory (NRAO) which is
operated by Associated Universities, Inc., under cooperative agreement
with the National Science Foundation. JLP is supported by an
appointment to the NASA Postdoctoral Program at the Jet Propulsion
Laboratory, California Institute of Technology, administered by Oak
Ridge Associated Universities through a contract with NASA. AH, JPB
and DP are grateful to the Australian Research Council for financial
assistance during this project via the Linkage International scheme
(Australia-France Co-operation Fund in Astronomy). SK is supported in
part by the National Research Foundation of Korea (NRF) grant funded
by the Korean government (MEST) 2009-0062866. AH thanks Erik
Rosolowsky, Adam Leroy, Alberto Bolatto, Kaye Marion and Michael
Murphy for helpful discussions. We thank the referee for constructive
criticism that improved the analysis in this paper.


\label{sect:bibliography}
\bibliographystyle{mn2e}
\bibliography{lmc,smc,cohi,dust,mocs,obstools,mnemonic,software,survey}
\label{lastpage}

\end{document}